\documentclass[zpreprint,zbstdefault]{zeus_paper}
\usepackage[english]{babel}
\usepackage[T1]{fontenc}
\usepackage{supertabular}
\chardef\usc=95
\chardef\til=126
\catcode`\@=11 
\DeclareRobustCommand\xdotspace{\futurelet\@let@token\@xdotspace}
\def\@xdotspace{%
  \ifx\@let@token.\else
  \ifx\@let@token\bgroup.\else
  \ifx\@let@token\egroup.\else
  \ifx\@let@token\/.\else
  \ifx\@let@token\ .\else
  \ifx\@let@token~.\else
  \ifx\@let@token!.\else
  \ifx\@let@token,.\else
  \ifx\@let@token:.\else
  \ifx\@let@token;.\else
  \ifx\@let@token?.\else
  \ifx\@let@token/.\else
  \ifx\@let@token'.\else
  \ifx\@let@token).\else
  \ifx\@let@token-.\else
  \ifx\@let@token\@xobeysp.\else
  \ifx\@let@token\space.\else
  \ifx\@let@token\@sptoken.\else
   .\space
   \fi\fi\fi\fi\fi\fi\fi\fi\fi\fi\fi\fi\fi\fi\fi\fi\fi\fi}
\catcode`\@=12 

\newcommand{\stru}[2]{%
   \relax\ifmmode\hbox{\vrule height#1 depth#2 width0pt}%
   \else\vrule height#1 depth#2 width0pt\fi}

\newcommand{\Ronum}[1]{\uppercase\expandafter{\romannumeral#1}}
\newcommand{\ronum}[1]{\expandafter{\romannumeral#1}}
\DeclareRobustCommand{\LaTeXZ}{%
  \LaTeX\kern-.05em4\kern-.1em
  {\raisebox{-0.2ex}{$\scriptstyle\text{ZEUS}$}}\xspace}

\DeclareMathAlphabet{\mathbf}{OT1}{cmr}{bx}{sl}
\newcommand{\eVdist}{\kern-0.06667em}

\newcommand{\slashfrac}[2]{%
  \raisebox{0.5ex}{\ensuremath #1}\kern-0.12em/\kern-0.08em
  \raisebox{-.8ex}{\ensuremath #2}}

\newcommand{\sqr}[3]{%
    {\vcenter{\hrule height.#3ex\hbox{\vrule width.#2ex height#1ex
     \kern#1ex\vrule width.#3ex}\hrule height.#2ex}}}

\catcode`\@=11 
\newcommand{\parenbar}{\mathpalette\p@renb@r}
\def\p@renb@r#1#2{\vbox{%
  \ifx#1\scriptscriptstyle \dimen@.7em\dimen@ii.2em\else
  \ifx#1\scriptstyle \dimen@.8em\dimen@ii.25em\else
  \dimen@1em\dimen@ii.4em\fi\fi \offinterlineskip
  \ialign{\hfill##\hfill\cr
    \vbox{\hrule width\dimen@ii}\cr
    \noalign{\vskip-.3ex}%
    \hbox to\dimen@{$\mathchar300\hfil\mathchar301$}\cr
    \noalign{\vskip-.3ex}%
    $#1#2$\cr}}}
\catcode`\@=12 

\newcommand{\IP}{{\rm I$\kern-0.01667em$P}\xspace}

\mathchardef\qsm=63
\mathchardef\pls=43
\mathchardef\mns=512
\mathchardef\plm=518
\mathchardef\eql=61
\mathchardef\smallleft=300
\mathchardef\smallright=301
\mathchardef\les=316
\mathchardef\gre=318
\mathchardef\leq=532
\mathchardef\grq=533

\catcode`\@=11 
\newcounter{pict@width}
\newcounter{pict@height}
\newlength{\pict@scale}
\newcommand{\psfigadd}[4]{%
\setcounter{pict@width}{1*\ratio{#2+\pict@scale/2}{\pict@scale}}
\setcounter{pict@height}{1*\ratio{#3+\pict@scale/2}{\pict@scale}}
\setlength{\unitlength}{\pict@scale}
\hbox to #2{\hspace{-\fill}\begin{picture}(\thepict@width,\thepict@height)
\put(0,0){\psfig{figure=#1,width=#2,height=#3,clip=}}
\SetScale{0.283466457}
\SetWidth{1.763889}
{#4}
\end{picture}}
}
\newcounter{pict@widthfst}
\newcounter{pict@widthscd}
\newcounter{pict@widthtot}
\newcommand{\psfigaddtwo}[7]{%
\setcounter{pict@widthfst}{1*\ratio{#2+\pict@scale/2}{\pict@scale}}
\setcounter{pict@widthscd}{1*\ratio{#2+#4+\pict@scale/2}{\pict@scale}}
\setcounter{pict@widthtot}{1*\ratio{#2+#4+#6+\pict@scale/2}{\pict@scale}}
\setcounter{pict@height}{1*\ratio{#3+\pict@scale/2}{\pict@scale}}
\setlength{\unitlength}{\pict@scale}
\hbox{\hspace{-\fill}\begin{picture}(\thepict@widthtot,\thepict@height)
\put(0,0){\psfig{figure=#1,width=#2,height=#3,clip=}}
\put(\thepict@widthscd,0){\psfig{figure=#5,width=#6,height=#3,clip=}}
\SetScale{0.283466457}
\SetWidth{1.763889}
{#7}
\end{picture}}
}
\newcommand{\psfigror}[4]{%
\setcounter{pict@width}{1*\ratio{#2+\pict@scale/2}{\pict@scale}}
\setcounter{pict@height}{1*\ratio{#3+\pict@scale/2}{\pict@scale}}
\setlength{\unitlength}{\pict@scale}
\hbox{\begin{picture}(\thepict@width,\thepict@height)
\put(0,\thepict@height){\psfig{figure=#1,width=#3,height=#2,clip=,angle=270}}
\SetScale{0.283466457}
\SetWidth{1.763889}
{#4}
\end{picture}}
}
\newcommand{\psfigrol}[4]{%
\setcounter{pict@width}{1*\ratio{#2+\pict@scale/2}{\pict@scale}}
\setcounter{pict@height}{1*\ratio{#3+\pict@scale/2}{\pict@scale}}
\setlength{\unitlength}{\pict@scale}
\hbox{\begin{picture}(\thepict@width,\thepict@height)
\put(0,0){\psfig{figure=#1,width=#3,height=#2,clip=,angle=90}}
\SetScale{0.283466457}
\SetWidth{1.763889}
{#4}
\end{picture}}
}
\catcode`\@=12 
\newlength\listtextwidth

\catcode`\@=11 
\newlength{\@tabfninsert}
\newlength{\@tabfnwidth}
\newcommand{\tabfootnote}[2]{%
  \setlength{\@tabfninsert}{0.8em}
  \setlength{\@tabfnwidth}{\textwidth}
  \addtolength{\@tabfnwidth}{-\@tabfninsert}
  \addtolength{\@tabfnwidth}{-0.4em}
  \noindent\makebox[\@tabfninsert][r]{\footnotesize$^{#1}$\hfil}\hfill%
  \parbox[t]{\@tabfnwidth}{\footnotesize #2\hfill}}
\catcode`\@=12 

\newcommand{\CPC}[3]{Comp. Phys.\ Comm.\ {\bf#1}, #3 (#2)}

\newcommand{\NP}[3]{Nucl.\ Phys.\ {\bf#1}, #3 (#2)}
\newcommand{\NPPS}[3]{Nucl.\ Phys.\ Proc.\ Suppl.\ {\bf#1}, #3 (#2)}
\newcommand{\PL}[3]{Phys.\ Lett.\ {\bf#1}, #3 (#2)}

\newcommand{\PR}[3]{Phys.\ Rev.\ {\bf#1}, #3 (#2)}
\newcommand{\PRL}[3]{Phys.\ Rev.\ Lett.\ {\bf#1}, #3 (#2)}

\newcommand{\NIM}[3]{Nucl.\ Instr.\ and Meth.\ {\bf#1}, #3 (#2)}
\newcommand{\ZP}[3]{Z.\ Phys.\ {\bf#1}, #3 (#2)}
\newcommand{\EP}[3]{Eur.\ Phys.\ J.\ {\bf#1}, #3 (#2)}

\def\mx2{${ M_{x}^{2} } $}

\def\Q2{${ Q^{2} }  $}
\def\q2{${ Q^{2} }  $}

\def\e+p{$ e^+-p$ }

\def\ltap{\raisebox{-.4ex}{\rlap{$\sim$}} \raisebox{.4ex}{$<$}}
\def\gtap{\raisebox{-.4ex}{\rlap{$\sim$}} \raisebox{.4ex}{$>$}}

\begin{document}

\prepnum{DESY--02--142\\
September 2002}

\title{
Leading proton production \\
in \boldmath{$e^+p$} collisions at HERA}                                                       
                   
\author{ZEUS Collaboration}

\date{}

\confname{XXXIst International Conference on High Energy Physics}
\confplacedate{24--31 July 2002, Amsterdam, The Netherlands}
\confsession{Parallel Session(s): 6}
\confabsnum{833}

\abstract{
Events with a final-state proton carrying a large fraction of the
proton beam momentum, $x_L>0.6$, and the square of the transverse momentum
$p_T^2 <0.5$~GeV$^2$, have been studied in $e^+p$ collisions with the
ZEUS detector at HERA.
Data with different photon virtualities 
were used: $Q^2<0.02$~GeV$^2$, $0.1 <Q^2<0.74$~GeV$^2$ and 
$3<Q^2<254$~GeV$^2$, corresponding to integrated luminosities
of 0.9, 1.85 and 3.38~pb$^{-1}$.
The cross sections are given as a function of
$x_L$, $p_T^2$, $Q^2$ and the Bjorken scaling variable, $x$. 
The ratio of the cross section for leading proton production to the 
inclusive $e^+p$ cross section shows only a mild dependence on $Q^2$ and 
on $x$. 
In the  region $0.6<x_L<0.97$, the relative yield of protons
is only a weak function of $x_L$.
}

\makezeustitle 

\def\3{\ss}                                                                                        
\newcommand{\address}{ }                                                                           
\pagenumbering{Roman}                                                                              
                                                   %
\begin{center}                                                                                     
{                      \Large  The ZEUS Collaboration              }                               
\end{center}                                                                                       
  S.~Chekanov,                                                                                     
  D.~Krakauer,                                                                                     
  J.H.~Loizides$^{   1}$,                                                                          
  S.~Magill,                                                                                       
  B.~Musgrave,                                                                                     
  J.~Repond,                                                                                       
  R.~Yoshida\\                                                                                     
 {\it Argonne National Laboratory, Argonne, Illinois 60439-4815}~$^{n}$                            
\par \filbreak                                                                                     
  M.C.K.~Mattingly \\                                                                              
 {\it Andrews University, Berrien Springs, Michigan 49104-0380}                                    
\par \filbreak                                                                                     
  P.~Antonioli,                                                                                    
  G.~Anzivino$^{   2}$,                                                                            
  G.~Bari,                                                                                         
  M.~Basile,                                                                                       
  L.~Bellagamba,                                                                                   
  D.~Boscherini,                                                                                   
  A.~Bruni,                                                                                        
  G.~Bruni,                                                                                        
  G.~Cara~Romeo,                                                                                   
  M.~Chiarini,                                                                                     
  L.~Cifarelli,                                                                                    
  F.~Cindolo,                                                                                      
  A.~Contin,                                                                                       
  M.~Corradi,                                                                                      
  S.~De~Pasquale,                                                                                  
  P.~Giusti,                                                                                       
  G.~Iacobucci,                                                                                    
  G.~Levi,                                                                                         
  A.~Margotti,                                                                                     
  T.~Massam,                                                                                       
  R.~Nania,                                                                                        
  C. Nemoz$^{   3}$,                                                                               
  F.~Palmonari,                                                                                    
  A.~Pesci,                                                                                        
  \mbox{G.~Sartorelli},                                                                            
  Y.~Zamora Garcia$^{   4}$,                                                                       
  A.~Zichichi  \\                                                                                  
  {\it University and INFN Bologna, Bologna, Italy}~$^{e}$
% ...
\newpage                                                                                           
$^{\    1}$ also affiliated with University College London, UK \\                                  
$^{\    2}$ now at Universit\`a di Perugia, Dipartimento di Fisica, 
Perugia, Italy \\               
$^{\    3}$ now at E.S.R.F., Grenoble, France \\                                                    
$^{\    4}$ now at Inter American Development Bank,                                                
Washington DC, USA\\                                                                                
% ...
$^{\    6}$ on leave of absence at University of                                                   
Erlangen-N\"urnberg, Germany\\                                                                     
% ...
                                                           %
                                                           %
\newpage   
                                                           %
                                                           %
\begin{tabular}[h]{rp{14cm}}                                                                       
$^{a}$ &  supported by the Natural Sciences and Engineering Research                               
          Council of Canada (NSERC) \\                                                             
$^{b}$ &  supported by the German Federal Ministry for Education and                               
          Research (BMBF), under contract numbers HZ1GUA 2, HZ1GUB 0, HZ1PDA 5, HZ1VFA 5\\         
$^{c}$ &  supported by the MINERVA Gesellschaft f\"ur Forschung GmbH, the                          
          Israel Science Foundation, the U.S.-Israel Binational Science                            
          Foundation and the Benozyio Center                                                       
          for High Energy Physics\\                                                                
$^{d}$ &  supported by the German-Israeli Foundation and the Israel Science                        
          Foundation\\                                                                             
$^{e}$ &  supported by the Italian National Institute for Nuclear Physics (INFN) \\                
$^{f}$ &  supported by the Japanese Ministry of Education, Culture, Sports, Science                
          and Technology (MEXT) and its grants for Scientific Research\\                           
$^{g}$ &  supported by the Korean Ministry of Education and Korea Science                          
          and Engineering Foundation\\                                                             
$^{h}$ &  supported by the Netherlands Foundation for Research on Matter (FOM)\\                   
$^{i}$ &  supported by the Polish State Committee for Scientific Research,                         
          grant no. 620/E-77/SPUB-M/DESY/P-03/DZ 247/2000-2002\\                                   
$^{j}$ &  partially supported by the German Federal Ministry for Education                         
          and Research (BMBF)\\                                                                    
$^{k}$ &  supported by the Fund for Fundamental Research of Russian Ministry                       
          for Science and Edu\-cation and by the German Federal Ministry for                       
          Education and Research (BMBF)\\                                                          
$^{l}$ &  supported by the Spanish Ministry of Education and Science                               
          through funds provided by CICYT\\                                                        
$^{m}$ &  supported by the Particle Physics and Astronomy Research Council, UK\\                   
$^{n}$ &  supported by the US Department of Energy\\                                               
$^{o}$ &  supported by the US National Science Foundation\\                                        
$^{p}$ &  supported by the Polish State Committee for Scientific Research,                         
          grant no. 112/E-356/SPUB-M/DESY/P-03/DZ 301/2000-2002, 2 P03B 13922\\                    
$^{q}$ &  supported by the Polish State Committee for Scientific Research,                         
          grant no. 115/E-343/SPUB-M/DESY/P-03/DZ 121/2001-2002, 2 P03B 07022\\                    
\end{tabular}                                                                                      
                                                           %
                                                           %

\pagenumbering{arabic} 
\pagestyle{plain}

\section{Introduction}
Events with a final-state proton carrying a large fraction of the
available energy but a small transverse momentum have been studied in
detail in high-energy hadron-proton collisions~\cite{review1}. The cross
section for such leading proton events shows a peak for
values of the final-state proton momentum close to 
the maximum kinematically allowed value, the so-called diffractive peak. 
For lower momenta, the cross section is lower 
and the fraction of events with a leading proton is 
approximately independent of the energy and type 
of the incoming particle. This characteristic behaviour has 
led to studies of the associated event in terms of the 
effective energy available for hadronisation~\cite{review2, review2a}. 
More recently, events with neutrons or protons carrying a  
large fraction of the proton-beam momentum have also
been measured in positron-proton ($e^+p$) scattering at 
HERA~\cite{FNC,H1LP,zeuslndijet,h1php,neutrons}.

The study of semi-inclusive rates in hadron-nucleon collisions indicates 
that particle production from the target nucleon is independent of the
type
of the incident hadron, a property known as vertex
factorisation~\cite{review1}. This has been studied for instance by
comparing semi-inclusive rates, normalised to the respective total cross
sections, for the production of particles in the fragmentation
region of  the target nucleon. The hadronic data~\cite{ISR} also show
that, in the
high-energy limit, the momentum distribution of the particles 
from the fragmentation of the target hadron is independent of 
the energy of the incoming particle.
These characteristics have not yet been extensively studied for baryon
production in electron-proton collisions.

Electroproduction of leading baryons is also interesting in other 
respects. 
The virtual photon mediating the interaction, in the reference frame 
in which the proton is at rest, fluctuates into a vector-meson-like object 
(the vector dominance model, VDM~\cite{vdm}). The transverse size of 
this projectile can be varied
by changing the virtuality, $Q^2$, of the photon. Real photons ($Q^2=0$) have
hadronic size, while, as $Q^2$ increases, the photon size decreases. 
It is thus possible to experiment with a  
projectile of varying size. 
This may lead, for instance, to different absorptive rescattering 
of the produced baryon as $Q^2$ changes~\cite{dalesio}, and hence 
to 
a breaking of vertex factorisation. 

Several mechanisms have been suggested to explain the hadroproduction or 
electroproduction of leading protons. None of them are as yet amenable to
calculations based on perturbative quantum chromodynamics (pQCD).
 This is, in part, a consequence of the fact that the small $p_T$ 
of the leading proton necessitates  a non-perturbative approach. Some 
models~\cite{SULLIVAN,ZOLLER,HOLTMANN,KOPE,SNS} are based on the 
Regge formalism, with leading proton production occurring through 
$t$-channel exchanges, both isoscalar and isovector, notably Pomerons, 
Reggeons and pions. These exchanges mediate the interaction between the 
proton and the fluctuations of the virtual photon. The relative 
contribution of the different exchanges varies as a function 
of the momentum and type of the scattered baryon: for leading 
protons, Pomeron exchange dominates in the diffractive-peak region 
with  Reggeon and pion exchanges contributing 
for lower outgoing-proton momenta. Other theoretical models 
retain quarks and gluons as fundamental entities, but add non-perturbative 
elements, such as soft-colour interactions~\cite{sci}. 
The concept of fracture functions also offers a general theoretical 
framework for a QCD-based study of leading baryon physics~\cite{FF}.

This paper reports studies of leading proton production in $e^+p$ 
collisions, 
$e^+p \rightarrow e^+Xp$, emphasizing the non-diffractive region. This
complements the recent ZEUS study of leading 
neutrons~\cite{neutrons}. High-energy protons with low transverse
momentum carrying at least 60\% of the incoming-proton momentum 
were measured in the ZEUS leading proton spectrometer (LPS)~\cite{lpsrho}.
The fraction of such events with a large rapidity gap in the forward
region is presented. The 
longitudinal- and transverse-momentum spectra are studied for different 
photon virtualities, from quasi-real photoproduction ($Q^2 \ltap
0.02$~GeV$^2$) to $Q^2 =254$~GeV$^2$. 
The dependence of the cross section for the production of leading
protons on the deep inelastic scattering (DIS) variables $x$ 
and $Q^2$ is also measured and 
compared to that for the inclusive reaction $e^+p \rightarrow
e^+X$. The results are discussed in the context of vertex factorisation
and particle-exchange models.
Finally, the properties of events with a leading proton and two jets are
presented.

\section{Experimental set-up}
\label{sec:setup}

The measurements were performed at the DESY $ep$ collider HERA using the 
ZEUS detector. In 1994 and 1995, HERA operated at a proton energy 
$E_p= 820$~GeV and a positron energy $E_e=27.5$~GeV.

A detailed description of the ZEUS detector can be found
elsewhere~\cite{bluebook}.
A brief outline of the components
that  are most relevant for this analysis is given below.

Charged particles are tracked by the central tracking detector (CTD),
which operates in a magnetic field of 1.43\,T provided
by a thin superconducting coil.
The CTD consists of 72 cylindrical drift-chamber layers,
organised in nine superlayers covering the polar-angle\footnote{
The ZEUS coordinate system is a right-handed Cartesian system,
with the $Z$ axis pointing in the proton beam direction,
referred to as the ``forward direction'',
and the $X$ axis pointing left towards the centre of HERA.
The coordinate origin is at the nominal interaction point.
The pseudorapidity is defined as $\eta = -\ln (\tan \frac{\theta}{2})$,
where the polar angle, $\theta$, is measured
with respect to the proton beam direction.}
region \mbox{$15^\circ < \theta < 164^\circ$.}
The relative transverse-momentum resolution for full-length tracks
is $\sigma(p_t)/p_t=0.0058p_t\oplus 0.0065 \oplus 0.0014/p_t$,
with $p_t$ in GeV \cite{ctd}.

The high-resolution uranium-scintillator calorimeter (CAL)~\cite{cal}
consists of three parts:
the forward (FCAL),  barrel (BCAL) and  rear (RCAL) calorimeters.
Each part is subdivided transversely into towers
and longitudinally into one electromagnetic section (EMC)
and either one (in RCAL) or two (in FCAL and BCAL)
hadronic sections (HAC).
The relative CAL energy resolutions are $\sigma(E)/E=0.18/\sqrt{E}$ for
electrons
and $\sigma(E)/E=0.35/\sqrt{E}$ for hadrons ($E$ in GeV). 

A lead-scintillator calorimeter (LUMI-$e$) at $Z=-35$~m~\cite{lumi}, with
a relative energy resolution of $\sigma(E)/E=0.18/\sqrt{E}$ ($E$ in GeV), 
was used to tag events with positrons scattered through
angles up to about 5~mrad, 
and to measure the scattered-positron energy, $E_e^{\prime}$, over the 
range $7<E_e^{\prime}<21$~GeV. These events have $Q^2 <0.02$~GeV$^2$ and are 
hereafter referred to as the ``photoproduction'' sample. This 
sample was collected in 1994 and corresponds to an integrated luminosity 
of $0.898\pm 0.014$~pb$^{-1}$.

Low-$Q^2$ events  ($0.1 < Q^2 < 0.74$~GeV$^2$) were tagged by requiring
the identification of the scattered positrons in the beam pipe calorimeter 
(BPC) \cite{zeusbpc,zeusbpt,surrow,monteiro}, a tungsten-scintillator 
sampling 
calorimeter, located close to the beam pipe, 3\,m downstream of the 
interaction point in the positron beam direction. 
This low-$Q^2$ sample, hereafter referred to as the ``BPC sample'', has
an integrated luminosity of $1.85\pm0.02$~pb$^{-1}$ and was 
collected in 1995.

For higher-$Q^2$ events ($Q^2>3$~GeV$^2$), the impact position 
on the CAL surface of the scattered
positron was determined with the small-angle rear tracking detector
(SRTD)~\cite{srtd} or the CAL. The SRTD is attached to the front face of 
the RCAL and consists of two planes of scintillator strips, 1~cm wide and 
0.5~cm thick, arranged in orthogonal orientations.
Events with 
$Q^2>3$~GeV$^2$ are referred to as the ``DIS sample'' in the following.
The 
integrated luminosity of this sample, which was collected in 1995, 
is $3.38\pm 0.03$~pb$^{-1}$.

The leading proton spectrometer (LPS)~\cite{lpsrho} detected
charged particles scattered at small angles and carrying a substantial 
fraction 
of the incoming-proton momentum; these particles remain 
in the beam pipe and their trajectories were measured by a system of 
silicon micro-strip detectors inserted very close (typically a few mm) 
to the proton beam. The detectors were grouped in six stations, 
S1 to S6, placed along the beam-line in the direction of the  
proton beam, between $Z=20$~m and $Z=90$~m.
The track deflections induced by the magnets in the 
proton beam-line allow a momentum analysis of the scattered proton.
During data taking, the detector planes were inserted close to the beam
by means of re-entrant  
pots and were retracted during beam dump and 
fill operations of the HERA machine.
For the present measurements, only the stations S4, S5 and S6 were used.
The intrinsic resolution is better than $1\%$ on the longitudinal momentum 
and 5~MeV on the transverse momentum. 
The effective transverse-momentum resolution is, however, dominated 
by the intrinsic transverse-momentum spread of the proton beam at the 
interaction point, which was $\approx 40$~MeV in the horizontal plane and 
$\approx 100$~MeV in the vertical plane.

\section{Kinematics and cross sections}

\begin{figure}[htb]
\begin{center}
\leavevmode
\hbox{%
\epsfxsize = 10cm
\epsffile{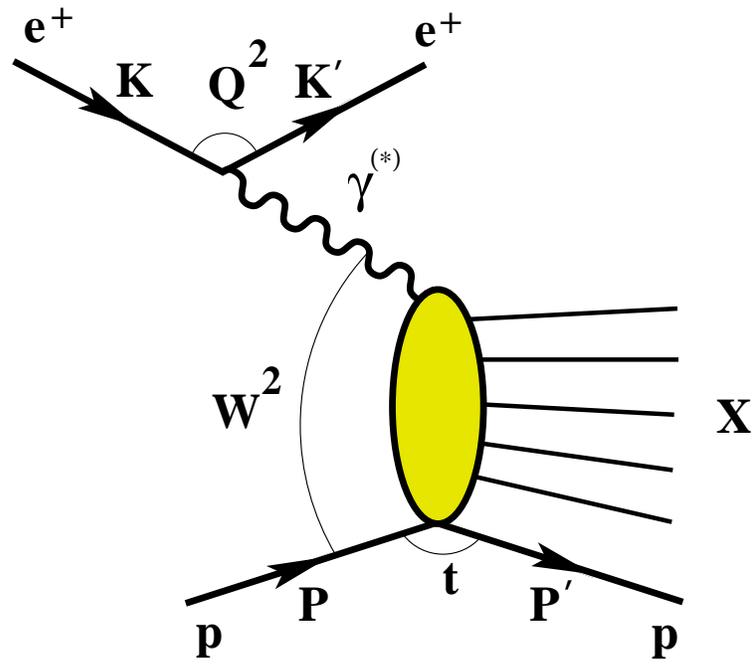}}
\end{center}
\vspace{4cm}
\caption{{\protect{Schematic diagram of the reaction $e^+p \rightarrow
e^+Xp$.
}}}
\label{fig:fig1}
\end{figure}

Figure~\ref{fig:fig1} illustrates 
semi-inclusive leading proton production in $ep$ collisions.
Four kinematic variables are needed to describe the interaction
$e^+ p\rightarrow e^+ X p$.
They are defined in terms of the four-momenta of the incoming and
outgoing positron, $K$ and $K'$, and of the incoming and outgoing
proton, $P$ and $P'$, respectively.

Two of the kinematic variables were chosen from among the Lorentz invariants 
used in inclusive studies, of which only two are independent:
$Q^2= -q^2 = -(K-K')^2$, the  virtuality 
of the exchanged photon;
$x=Q^2/(2P\cdot q)$ and $y=q\cdot P/(K\cdot P)\simeq Q^2/(sx)$;
and $W^2=(P+K-K')^2=m^2_p+Q^2(1-x)/x$, the square of the photon-proton
centre-of-mass energy.
In these equations, $m_p$ is the mass of the proton
and $\sqrt{s}=$ 300~GeV is the $e^+p$ centre-of-mass energy.

Two additional variables are required to describe the leading proton. They
can be chosen as the momentum fraction carried by the outgoing proton
\[
x_L =\frac{P' \cdot K}{P\cdot K}
\]
and its transverse momentum with respect to the direction of the 
incoming proton, $p_T$. In terms of these variables, the square of the
four-momentum
transfer from the target proton is given by
\begin{equation}
t=(P-P')^2 \simeq -\frac{p_T^2}{x_L} - \frac{(1-x_L )^2}{x_L 
}m^2_p, \nonumber
\end{equation}
where the second term is the minimum kinematically allowed value of $|t|$
for a given $x_L $. In a particle-exchange model, $t$ is the square 
of the four-momentum of the exchanged particle. 
The $p_T^2$ range covered by the
present data, $0<p_T^2<0.5$~GeV$^2$, thus translates into
$0<|t|<0.5$~GeV$^2$ 
at $x_L=1$ and 0.2~$\ltap$~$|t|$~$\ltap$~1~GeV$^2$ at $x_L=0.6$.          

The differential cross section for  inclusive $e^+p\rightarrow e^+ X$
scattering is written in terms of the proton structure function,
$F_2(x,Q^2)$, as
\begin{equation}
\frac{d^2\sigma_{e^+p\rightarrow e^+ X} }{dx dQ^2} =
 \frac{4\pi \alpha^2}{x Q^4}\left( 1-y+\frac{y^2}{2} \right)
  F_2(x , Q^2)(1+\Delta), 
  \label{eq:inc}
  \end{equation}
  where $\Delta$ is a correction that  takes
account of photon radiation, $Z^0$ exchange,
and the longitudinal structure function, $F_L$.
In analogy with this, the differential
cross section for semi-inclusive leading proton production,
$e^+p \rightarrow e^+ X p$, is written as
\begin{equation}
\frac{d^4\sigma_{e^+p\rightarrow e^+ Xp}}{dx dQ^2dx_L dp_T^2} =
 \frac{4\pi \alpha^2}{x Q^4}\left( 1-y+\frac{y^2}{2} \right)
   F^{\mbox{\rm\tiny LP(4)}}_2(x , Q^2, x_L, p_T^2)
     (1+\Delta_{LP}),   
\label{eq:ln}
\end{equation}

\noindent
where $\Delta_{LP}$ is the analogue of $\Delta$.


\subsection{Reconstruction of the kinematic variables}

Three samples of data were used: 
\begin{itemize}
\item the photoproduction sample, with the scattered positron 
tagged in the LUMI-$e$ calorimeter; 
\item the BPC sample, with the scattered positron measured in the BPC;
\item the DIS sample,  
with  the scattered positron  detected in the CAL. 
\end{itemize}

Different methods were used for the reconstruction of 
the kinematic variables, $Q^2$ and $W$, for the three data sets. 
Tagging of the scattered positron in the LUMI-$e$ calorimeter 
for photoproduction events does not allow the measurement of  
$Q^2$ event by event; however, the angular acceptance of
the LUMI-$e$ calorimeter limits the $Q^2$ range to the region 
$Q^2< 0.02$~GeV$^2$. For these events, $W$ was measured from $W^2 = ys$,
with $y=(E_e-E_e^{\prime})/E_e$, where $E_e'$ denotes the
energy of the outgoing positron. For the BPC sample, $E_e$ and the
positron scattering angle, $\vartheta_e$, as measured in the BPC, were
used (``electron method''~\cite{zeusbpc}) to
determine $Q^2$, $W$, $x$ and $y$. For the DIS sample, these variables 
were reconstructed using the double angle method~\cite{DA}.

For the reconstruction of the hadronic final state, $X$, 
the energy deposits in the CAL and the track momenta measured in the CTD
for the charged particles 
were clustered into  energy-flow objects (EFOs) which are assumed to
correspond to particles and are assigned the pion 
mass~\cite{zeusdiff94,gb}. The EFOs were used to reconstruct the mass, 
$M_X$, of the hadronic final state contained in the central detector. 
Using the EFOs, the $y$ variable was also reconstructed with the 
Jacquet-Blondel method \cite{jacblo}, which uses information from the 
hadronic final state to reconstruct the event kinematics,
and was denoted by $y_{JB}$. Furthermore, the variable
\begin{equation}
   \delta=\sum_{i}(E_i-p_{Z,i})+E_{e}'(1-\cos\vartheta_e) \nonumber
   \label{rec-empz}
\end{equation}
was evaluated, where $\sum_i$ denotes a sum over
all EFOs, excluding those assigned to the scattered positron, and $E_i$
and
$p_{Z,i}$ are the energy and
the longitudinal momentum of each EFO, respectively. 
For perfect resolution and fully contained
events, energy and momentum conservation constrain $\delta$ to be twice the 
positron beam energy.
The angle of the hadronic 
final state (as measured in the ZEUS central detector) with respect to the 
incoming-proton direction was evaluated from

\[ 
\cos \gamma_h = \frac{ (\sum_{i} p_{T,i})^2 -(\sum_i (E_i-p_{z,i}))^2} 
                     { (\sum_{i} p_{T,i})^2 +(\sum_i (E_i-p_{z,i}))^2 },
\]
where the sums $\sum_i$ run over all EFOs excluding those assigned to the 
scattered positron.

The modulus of the momentum of the scattered proton, $p'$, was
measured in the LPS, along with its component perpendicular 
to the mean proton beam direction, $p_T$. The variable 
$x_L$ was evaluated as $x_L=p'/E_p$. The mean direction of 
the incoming proton beam was determined for each proton fill of HERA using 
the reaction $ep \rightarrow e \rho^0 p$  at  $Q^2 \approx
0$~\cite{lpsrho}. 

In the following, the term ``leading proton'' is used to indicate a 
positively charged particle detected in the LPS. 
In the present measurement, $x_L$ is restricted to values larger than
0.6.
Charged-particle production measured at the
ISR~\cite{review2,piontoproton} shows that
the pion-to-proton ratio at $x_L=0.6$ is about 10$\%$ and falls
rapidly for increasing values of $x_L$.

\section{Event selection}

Photoproduction events were selected at the trigger level by requiring a
coincidence between an energy deposit of at least 5~GeV in the LUMI-$e$
and of at least 464 MeV in the RCAL (excluding the towers immediately
adjacent to the beam-pipe) or at least 1250 MeV (including those towers).
This requirement helps to suppress the background from bremsstrahlung
events ($ep \rightarrow e \gamma p$) characterised by having a scattered
positron in the LUMI-$e$ and no activity in the rest of the detector.
The BPC and DIS events were triggered by requiring the presence of a
scattered positron in the BPC and the CAL, respectively.  No requirement
was imposed on the final-state proton at the trigger level.

Events were selected offline in three steps: first inclusive events with
the scattered positron in the LUMI-$e$, the BPC or the CAL were selected;
then, a track in the LPS was required; finally a search for jets was
carried out in the hadronic final state, X. The details of the selection
procedure are presented in the rest of this section. The following 
requirements were imposed for all samples:

\begin{itemize}

\item the $Z$ coordinate of the reconstructed vertex, if measured,  
in the range $-50 < Z < 50$~cm;

\item{} the timing of the interaction, as measured by the CAL, 
consistent with the timing of an $e^+p$ bunch crossing.

\end{itemize}

\noindent

For the BPC and DIS samples, the requirement $35 <\delta <65$~GeV was also 
imposed in order to reduce the photoproduction background and
to minimise the effect of the radiative corrections.

The photoproduction sample~\cite{tesi_yuri} was selected by requiring
that a  
positron be measured in the LUMI-$e$ with energy in the range 
$12<E_e^{\prime}<18$~GeV, corresponding to $176 <W< 225$~GeV. Overlay
events, in which some activity in the RCAL accidentally overlaps with the
scattered positron of a bremsstrahlung event ($ep \rightarrow e \gamma p$)
in the LUMI-$e$, were subtracted as discussed in an 
earlier study~\cite{t_lps}. The subtraction was less than 3\%.

The BPC sample~\cite{zeusbpc,tesi_alberto} was selected by requiring a
scattered positron measured in the BPC with $E_e^\prime > 7$~GeV and a 
photon virtuality in the range $0.1 <Q^2 <0.74 $~GeV$^2$. In addition,
the requirement $0.08<y<0.74$ was imposed, which corresponds to 
$85 < W < 258$~GeV. Finally, $y_{JB}>0.05$ was required, thus ensuring
hadronic activity away from the forward direction and reducing
the migration of events from low $y$, where the resolution of the electron 
method is poor.

In the DIS analysis\cite{tesi_alberto,tesi_yuri}, a scattered positron 
with energy $E_e^\prime>10$~GeV was required in the CAL; the photon 
virtuality was restricted to the
interval $3 <Q^2 <254$~GeV$^2$ and $W$ to the region $45 <W <225$~GeV.
Finally, the condition $y_{JB}>0.03$ was imposed. 

The total number of events thus selected was approximately 
94~000 for the photoproduction sample, 
50~000 for the BPC sample and 386~000 for the DIS sample.

Next, high-momentum protons in the LPS were selected 
by requiring:
\begin{itemize}

\item{} one track in the LPS with $p_T^2<0.5$~GeV$^2$ and 
$0.6<x_L<1.02$. For $x_L>0.97$, a lower bound on $p_T^2$ was also imposed: 
$p_T^2>0.073$~GeV$^2$. For the 1994 data,  the $p_T^2$ range was 
restricted to $p_T^2<0.04$~GeV$^2$. The $p_T^2$
cuts and the lower limit on $x_L$ restrict 
the data to a region of well understood acceptance;

\item{} 
no candidate track was accepted if, at any point, the minimum distance 
of approach to the beam pipe, $\Delta_{\rm pipe}$, was less than $0.4$~mm 
(0.5~mm for the 1994 data). This cut reduced the sensitivity of the acceptance 
to the uncertainty on the location of the beam-pipe apertures;

\item events in which the reconstructed proton 
track passed closer than a distance $\Delta_{\rm plane}=0.2$~mm 
to the edge of any LPS detector were rejected.  
This ensured that the track was well within the active region of 
the detectors;

\item{} the total $E+ p_Z$ of the event 
was required to be smaller than $1655$~GeV. 
For fully contained events, this quantity should be equal to 
$2 E_p =1640$~GeV.
This cut rejects random overlays of a beam-halo proton and an event
satisfying the trigger and selection cuts applied to the non-LPS
variables;

\item{} for $x_L>0.97$, $M_X>2$~GeV was required, where $M_X$ is 
the reconstructed hadronic mass in the CAL. This rejects
contributions from exclusive production  of low-mass vector mesons, 
which have $Q^2$ and $t$ dependences different from
those of the inclusive events~\cite{review_vm}. 

\end{itemize}

After this selection, the total number of events with a good LPS track was 
1834 for the photoproduction sample, 1697 for the BPC
sample and 13335 for the DIS sample.

Finally, a search was performed for jets in the hadronic final
state~\cite{tesi_mariacarmela}. 
Because of the limited statistics of the photoproduction and BPC samples,
the search was 
limited to the DIS data. The jets were reconstructed using 
the $k_T$ algorithm~\cite{JETKT}, requiring a jet transverse energy 
$E_T > 4~\mathrm{GeV}$ in the $\gamma^{\ast}p$ centre-of-mass system and a 
jet pseudorapidity, in the laboratory frame, in the range 
$-2 < \eta^{\rm jet} < 2.2$. A sample of 225 events with exactly two jets 
was selected.

\section{Monte Carlo simulation}

Several Monte Carlo (MC) generators were used to determine the acceptance
of the apparatus for events with a leading proton.
The EPSOFT2.0 Monte Carlo~\cite{mx,michal,thesis_masahide} was used for 
the BPC data. 
This generator simulates diffractive processes with dissociation of the 
virtual photon, as well as non-diffractive processes. Vertex factorisation
is assumed. 
The HERACLES4.6 Monte Carlo~\cite{heracles}, 
which simulates initial- and final-state QED 
radiation, is interfaced to EPSOFT. Samples of DIS events were simulated
with  
RAPGAP~\cite{rg,thesis_peter} version 2.06/06, which incorporates meson 
and Pomeron exchange; it also assumes vertex factorisation. QED radiation
was also simulated using HERACLES.
Weights were assigned to the events generated with EPSOFT and RAPGAP 
such that the reconstructed proton $x_L$ and $p_T^2$ spectra agreed
with the data. For the photoproduction data, and for systematic studies,
events that only contain a proton
with distributions in $x_L$ and
$p_T^2$ tuned to those of the data were generated. This simulation
produced the
same results for the LPS acceptance as EPSOFT and RAPGAP.

All generated events were passed through the trigger-simulation package and 
the standard ZEUS detector simulation, based on the GEANT 3.13 
program~\cite{geant}.
The simulation includes the geometry of the beam-pipe apertures,
the HERA magnets and their magnetic fields.
The spread of the interaction-vertex position was also simulated, as were
the proton-beam angle with respect to the nominal direction and its dispersion
at the interaction point. The simulated events were then passed through the 
same reconstruction and analysis programs as the data. 

\begin{figure}[htb]
\begin{center}
\leavevmode
\hbox{%
\epsfxsize = 16cm
\epsfysize = 16cm
\epsffile{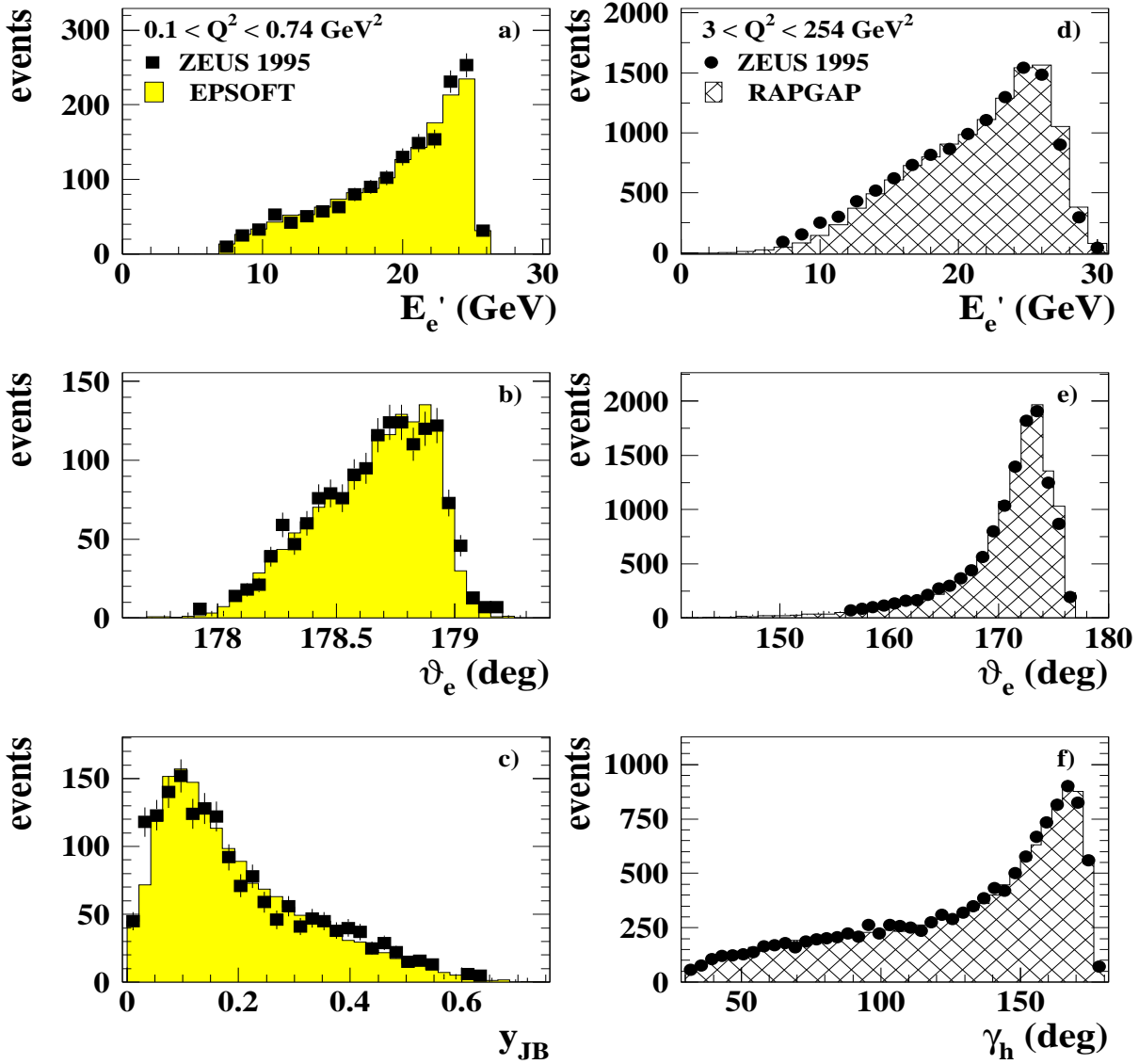}
}
\end{center}
\caption{
Distributions of the variables a) $E_e^{\prime}$, b) $\vartheta_e$  
and c) $y_{JB}$ for the reconstructed BPC data (squares) and the simulated 
events 
(EPSOFT), shown as the shaded histograms (normalised to the data); 
d) $E_e^{\prime}$, e) $\vartheta_e$ and
f) $\gamma_h$  for the reconstructed DIS data (dots) and the simulated 
events 
(RAPGAP), shown as the hatched histograms (normalised to the data). 
}
\label{data_mc_phys}
\end{figure}

Figures~\ref{data_mc_phys}a)-c) 
compare the distributions of the
reconstructed variables $E^{\prime}_e$, $\vartheta_e$, and $y_{JB}$  
in events generated with EPSOFT with those for the BPC data. The agreement
between the data and the simulated 
distributions is good. A similarly good description 
of the DIS data by RAPGAP is shown in Figs.~\ref{data_mc_phys}d)-f) for
the variables $E^{\prime}_e$, $\vartheta_e$ and $\gamma_h$.

\begin{figure}[htb]
\begin{center}
\leavevmode
\hbox{%
\epsfxsize = 16cm
\epsfysize = 16cm
\epsffile{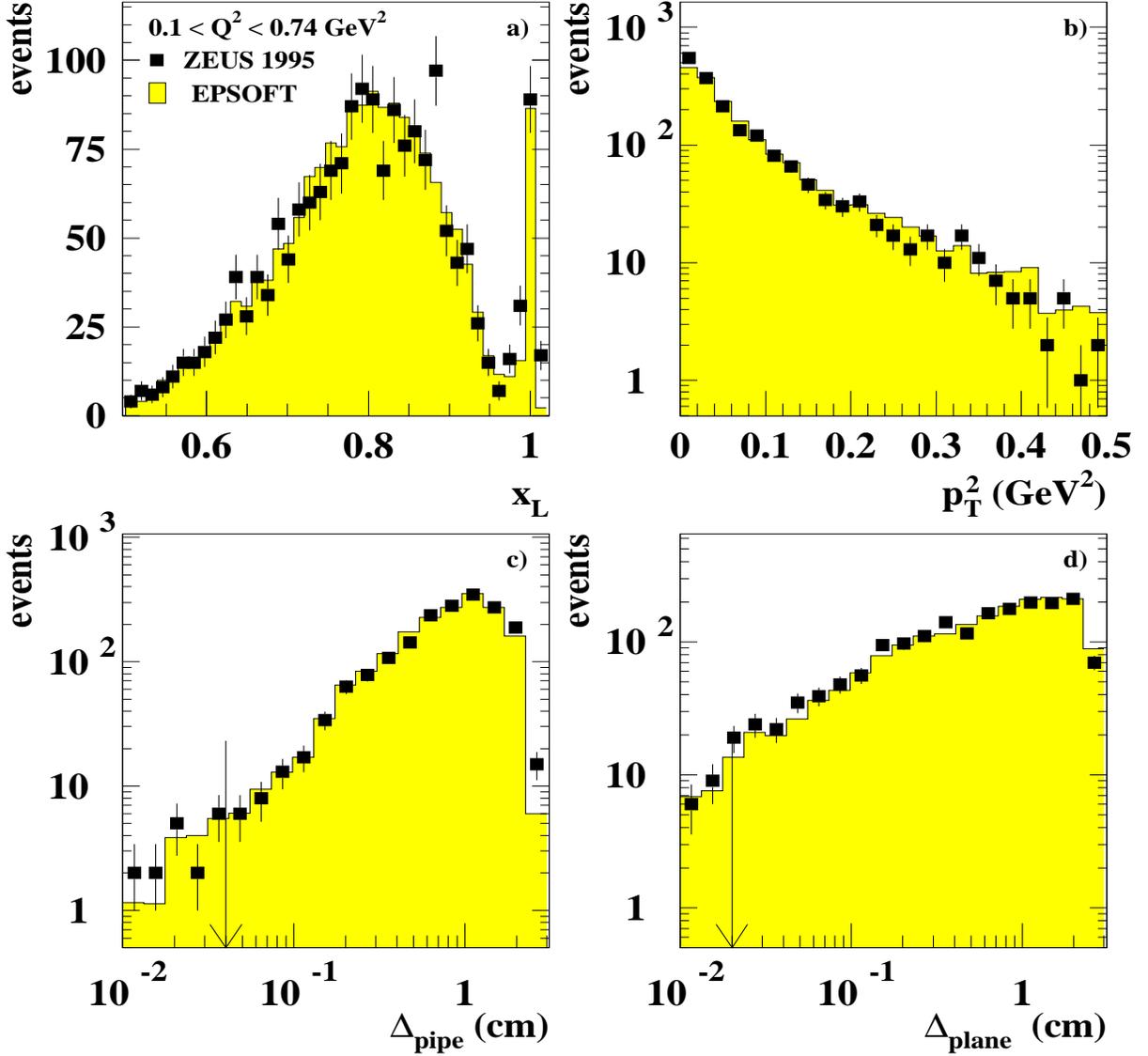}
}
\end{center}
\caption{Distributions of the variables a) $x_L$, b) $p_T^2$, 
c) $\Delta_{\rm pipe}$ and d) $\Delta_{\rm plane}$, for the reconstructed 
BPC data (squares) and for the simulated events (EPSOFT)
shown as the shaded histogram (normalised to the data). The arrows indicate
the minimum allowed  values of $\Delta_{\rm pipe}$ and $\Delta_{\rm
plane}$ (see text). 
}
\label{data_mc}
\end{figure}

The agreement between the data and the MC simulation of the leading-proton
variables is also good for all three samples. As an example, 
the distributions for the reconstructed EPSOFT
events as a function of $x_L$, $p_T^2$, $\Delta_{\rm pipe}$ and
$\Delta_{\rm plane}$
are compared with those of the BPC data in Fig.~\ref{data_mc}. 

\section{Acceptance}

\begin{figure}[htb]
\begin{center}
\leavevmode
\hbox{%
\epsfxsize = 14cm
\epsfysize = 20cm
\epsffile{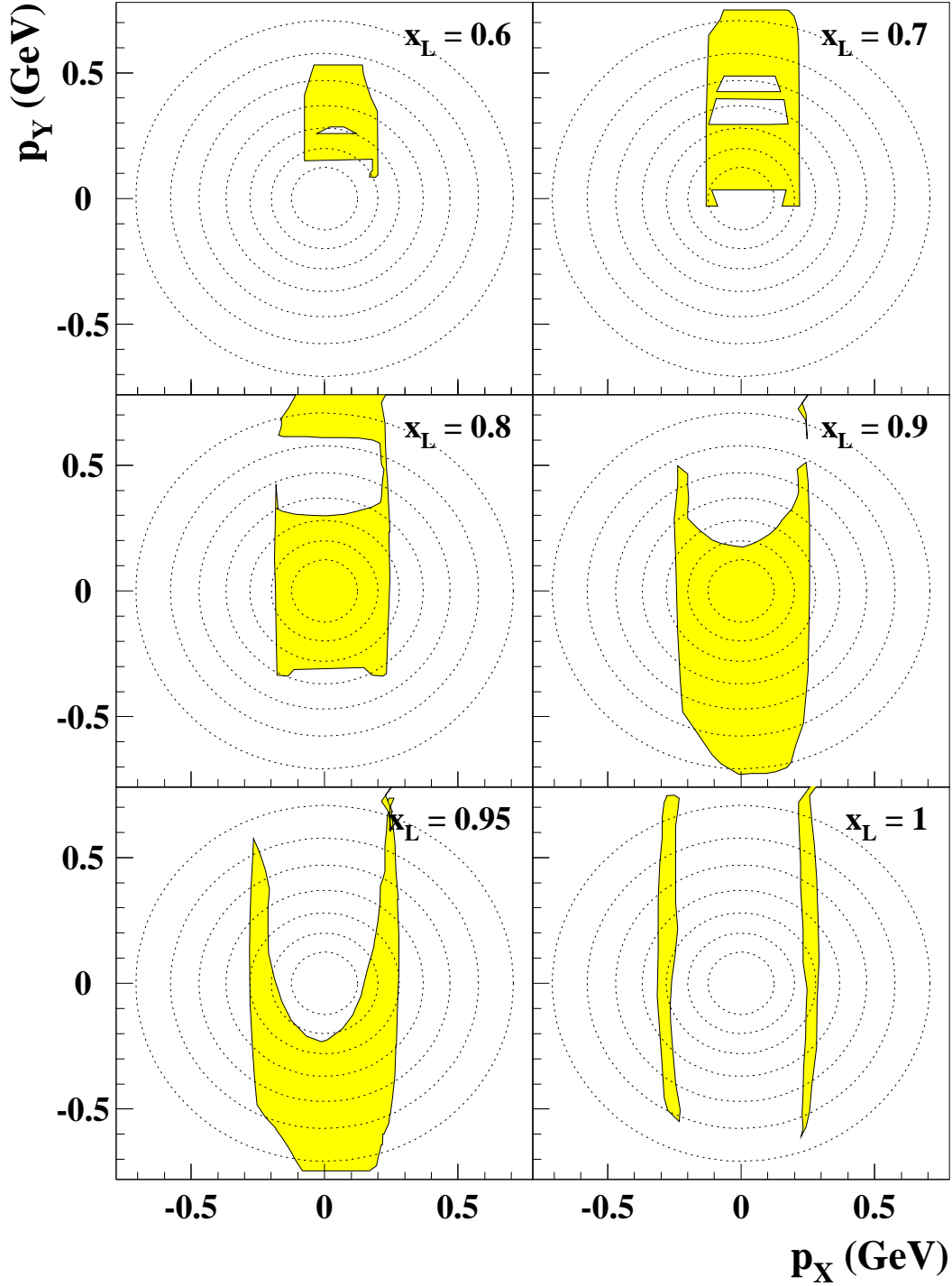}
}
\end{center}
\caption{{\protect{LPS geometrical acceptance for different $x_L$ values 
as a function of $p_X$ and $p_Y$. The shaded areas indicate the regions 
of acceptance. The dashed circles indicate the limits of the $p_T$ 
bins used in the analysis (the bin edges are 
$0.124~\mathrm{GeV}, 
 0.199~\mathrm{GeV},
 0.280~\mathrm{GeV}, 0.370~\mathrm{GeV}, 0.470~\mathrm{GeV}, 
0.507~\mathrm{GeV}, 0.707~\mathrm{GeV}$).}}}
\label{lps_acceptance}
\end{figure}

Figure~\ref{lps_acceptance} shows the ranges of  $p_X$ and $p_Y$
accessed by the LPS for six values of $x_L$, using the coincidence of any
two of the
S4, S5 and S6 stations. Here, $p_X$ and $p_Y$ are  the $X$ and $Y$
components of the scattered-proton momentum. The region covered is
determined by the beam-pipe apertures, the shape of the sensitive region
of the LPS detectors and the magnet strengths; it is limited
to $x_L~\gtap~0.5$ and $p_T^2=p_X^2+p_Y^2 ~\ltap~0.7$~GeV$^2$. Integrated
over the falling $p_T^2$ distribution, the LPS geometrical acceptance
reaches
a maximum for $x_L$ in the range 0.8 to 0.9. 

The acceptance was computed as the ratio of the number of reconstructed 
events in a bin of a given variable to the number of generated 
events in that bin. The acceptance thus includes the effects of the
geometrical 
acceptance of the apparatus, its efficiency and resolution, as well as the 
event selection and reconstruction efficiencies. Values of the acceptance
are given in Section~\ref{ratio-method}.

\section{Systematic uncertainties}
\label{sys_errors}

The systematic uncertainties 
were obtained by modifying the requirements and the analysis procedures as 
listed below:

\begin{itemize}

\item the sensitivity to the selection of the proton track was 
studied by the following procedure~\cite{lpsrho}:

\begin{itemize}

\item the track-selection requirements were varied. In particular, the 
minimum-allowed values of $\Delta_{\rm pipe}$ were changed from 0.2
to 0.6~mm and the minimum value of $\Delta_{\rm plane}$ was varied from
0.1 to 0.3~mm; 

\item the positions of some of the elements of the proton beam-line
were varied
within their uncertainties. This is particularly relevant at
low $x_L$, where the proton momentum is significantly lower than that of
the $x_L \approx 1$ protons used in the LPS alignment procedure;

\item the LPS detector positions varied slightly from fill to fill in the
94 sample. The small deviations of the acceptance implied by these
movements were estimated  by dividing the data into a ``low
acceptance'' and a ``high acceptance'' sample, depending on the positions
of
the LPS stations;

\end{itemize}

\item the sensitivity to the remaining selection cuts  
was also investigated:

\begin{itemize}

\item for the inclusive photoproduction sample, the selection cuts were tightened:
the $E_e'$ range was restricted to $13<E_e'<16$~GeV, corresponding to  
$195<W<215$~GeV and 
the minimum energy deposition in the RCAL was raised to 2 GeV. In addition, 
the correction for the bremsstrahlung background was 
removed~\cite{t_lps}; 

\item for the BPC sample, the BPC energy scale was varied by $\pm1\%$ and 
the sizes of the parameters in the 
BPC energy calibration were changed within 
their uncertainties~\cite{zeusbpc}. The selection limits on the 
positron-candidate 
shower width were also varied~\cite{zeusbpc};

\item for the DIS sample, the positron-selection procedure was varied.
The cut on the scattered-positron energy was changed to 8~GeV and
12~GeV and the size of the fiducial region for the impact 
position of the scattered positron in the SRTD was raised by
$\pm 0.5$~cm in both $X$ and $Y$;

\item for both the BPC and the DIS samples, the lower limit on $\delta$ was 
varied between 32 and 38~GeV and the upper limit between 60 and 68~GeV;
the cut on $y_{JB}$ was varied by $\pm 0.01$;

\item the allowed range of values for the $Z$ coordinate of the vertex
was restricted to $-40<Z<40$~cm. The effect of removing the vertex requirement 
was also studied;

\item in addition, for the jet studies, the minimum  jet energy was varied
between 3.8 and 4.2 GeV, and the upper limit on $\eta^{\rm jet}$ was 
varied between 2 and 2.4.

\end{itemize}
 
\end{itemize}

\noindent 
The total systematic uncertainty on the cross sections, obtained by 
summing all the above contributions in quadrature,  totalled about
$\pm$20\% at  $x_L \approx 0.65$, decreasing to $\pm$(10-15)\% for
$x_L~\gtap~0.75$. The dominant contributions are those related to  the 
track-selection requirements in the LPS.

\section{The ratio method}
\label{ratio-method}


%


In the following, several results are presented in terms of the ratios,
$r^{\rm LP(2)}$ and $r^{\rm LP(3)}$, of the cross section for production 
of leading protons to the cross section for inclusive $e^+p$ scattering;
these ratios are evaluated 
in bins of $x$ and $Q^2$ ($r^{\rm LP(2)}$), or in 
bins of $x$, $Q^2$ and $x_L$ ($r^{\rm LP(3)}$). They are obtained 
from the 
measured fraction of the events, in a given bin, that have a leading 
proton, $N^{\rm LP}/N$. In this fraction,
the acceptance corrections related to the positron selection procedure
cancel, and so do the corresponding systematic uncertainties. The only
remaining correction to apply is that for the LPS acceptance,
$\epsilon_{LPS}$.

The ratio $r^{\rm LP(2)}$ was thus obtained as
\begin{equation}
r^{\rm LP(2)}(x,Q^2) =
            \frac{N^{\rm LP}(x,Q^2)}{N(x,Q^2)}
                      \frac{1}{\epsilon_{LPS}} \, \nonumber .
\label{rlp2}
\end{equation}

\noindent
Averaged over the region 
$0.6<x_L<0.97$ and $p_T^2<0.5$~GeV$^2$, $\epsilon_{LPS}$ is approximately
23\%; over the region $0.6<x_L<0.97$ and $p_T^2<0.04$~GeV$^2$,
$\epsilon_{LPS} \approx 51\%$.

From Eqs.~(\ref{eq:inc}) and ~(\ref{eq:ln}), it is apparent that the cross 
section ratio $r^{\rm LP(2)}$ is also equal to the ratio of the proton 
tagged and inclusive structure functions:

\begin{equation}
r^{\rm LP(2)}(x,Q^2) 
                 = \frac{\bar{F}_2^{~\rm LP(2)}(x,Q^2)}
                            {F_2(x,Q^2)},
\label{rlp2a}
\end{equation}

\noindent where $\bar{F}_2^{\rm LP(2)}$ is obtained from $F_2^{\rm LP(4)}$
by integration over the measured $x_L$ and $p_T^2$ ranges:

\begin{equation}
\bar{F}_2^{\rm LP(2)}(x,Q^2) = 
    \int_0^{p_{T \max}^2} dp_T^2
    \int_{0.6}^{0.97} d x_L~F_2^{\rm LP(4)}(x,Q^2,x_L,p_T^2). \nonumber
\label{f2bar}
\end{equation}

The radiative 
corrections and the contributions from $F_L$ are assumed to be the same 
for 
the inclusive and the proton-tagged reactions.

The ratio $r^{\rm LP(3)}(x,Q^2, x_L)$ is defined in analogy to 
$r^{\rm LP(2)}(x,Q^2)$:

\begin{equation}
r^{\rm LP(3)}(x,Q^2, x_L) =
   \frac{N^{\rm LP}(x,Q^2, x_L)}{N(x,Q^2)}
                      \frac{1}{\epsilon_{LPS}(x_L) \Delta x_L}, 
\label{eq:rlp3}
\end{equation}

\noindent
where $\Delta x_L$ indicates the size of the $x_L$ bins. In analogy with
Eq.~(\ref{rlp2a}), 

\begin{equation}
r^{\rm LP(3)}(x,Q^2, x_L) = 
\frac{\bar{F}_2^{\rm LP(3)}(x,Q^2,x_L)}
                            {F_2(x,Q^2)}, 
\label{rlp3a}
\end{equation}

\noindent
where $\bar{F}_2^{\rm LP(3)}(x,Q^2,x_L)$
differs from $\bar{F}_2^{\rm LP(2)}(x,Q^2)$
in that no integration over $x_L$ is performed.

The ratios $r^{\rm LP(2)}$ and $r^{\rm LP(3)}$ can also be interpreted
in terms of the virtual photon-proton cross section for
the process $\gamma^* p \rightarrow Xp$ and the total virtual
photon-proton cross section, $\sigma_{\rm tot}$. For example,
the ratio $r^{\rm LP(3)}$ can be written as

\begin{equation}
r^{\rm LP(3)}(x,Q^2, x_L) = \frac{1}{\sigma_{\rm tot}}
\int_0^{p_{T \max}^2} d p_T^2 
\frac{d^2\sigma_{\gamma^{\ast} p \rightarrow Xp}} {dx_L dp_T^2}=
\frac{1}{\sigma_{\rm tot}}\frac{d\sigma_{\gamma^{\ast}p\rightarrow Xp}}{dx_L}, 
\nonumber
\label{rlp3sigtot}
\end{equation}

\noindent
where the virtual photon-proton cross section,
$d^2 \sigma_{\gamma^{\ast} p\rightarrow Xp}/dx_L dp_T^2$, is related to the 
positron-proton cross section, 
$d^4\sigma_{e^+ p\rightarrow e^+ Xp}/dQ^2 dx dx_L dp_T^2$, by

$$\frac{d^4\sigma_{e^+ p\rightarrow e^+ Xp}}{dQ^2 dx dx_L dp_T^2} =
\Gamma 
\frac{d^2 \sigma_{\gamma^{\ast} p\rightarrow Xp}}{dx_L dp_T^2},$$

\noindent
where $\Gamma= (\alpha/x Q^2 \pi) [ 1+(1-y^2)]$, is the photon flux
factor and 
$\alpha$ is the fine-structure constant.

\section{Models}
\label{models}

The data were tested against the hypothesis of vertex factorisation, 
a very general feature of hadron-hadron interactions~\cite{review1} which
relates reactions with different beam particles
to their respective total cross sections. In particular, in the reaction 
$\gamma^{\ast} p \rightarrow Xp$, the $\gamma^{\ast}$-$X$ and $p$-$p$ 
vertices factorise if the amplitude for
the reaction can be written as the product of two vertex functions,
$G_{\gamma^{\ast} X}(x,Q^2)$ and $G_{pp}(x_L,p_T^2)$. In this case, the
cross section as a function of the lepton variables $x$ and
$Q^2$ should be independent of the baryon variables $x_L$ and $p_T^2$, and
{\it vice versa}.

The data were also compared to the following specific models:

\begin{itemize}

\item the LUND string-fragmentation model as implemented in 
JETSET~\cite{jetset} and used in DJANGO~\cite{django}, in which leading 
baryons originate from the hadronisation  of the target;

\item the soft-colour-interaction model (SCI)~\cite{sci}, as
implemented in LEPTO 6.5 \cite{lepto}, where leading protons 
are obtained from standard DIS events by means of a 
non-perturbative redistribution of colour among the fragmenting partons;

\item the Regge model of Szczurek et al.~\cite{SNS}, which 
assumes a superposition of Pomeron, Reggeon and pion exchanges;

\item the QCD-inspired model of Dur\~aes et al.~\cite{igm}, developed in
the
framework of the interacting-gluon model~\cite{igm1}, which assumes that 
high-energy hadron-hadron collisions are dominated by multiple incoherent 
gluon-gluon interactions. The valence quarks 
that do not take part in the interaction give rise to the leading baryons. The
extension to $ep$ collisions is made in the VDM framework.
\end{itemize}

\section{Results}

\subsection{Leading proton events with a forward large rapidity gap}
\label{etamax}

Some indication of the production mechanism of leading protons can be
obtained from the rapidity distribution of the hadronic final-state
particles. In particular, events of diffractive origin, i.e. due to
Pomeron exchange, are characterised by a gap in the rapidity distribution
in the forward direction.

\begin{figure}[htb]
\begin{center}
\leavevmode
\hbox{%
\epsfxsize=16cm
\epsfysize=16cm
\epsffile{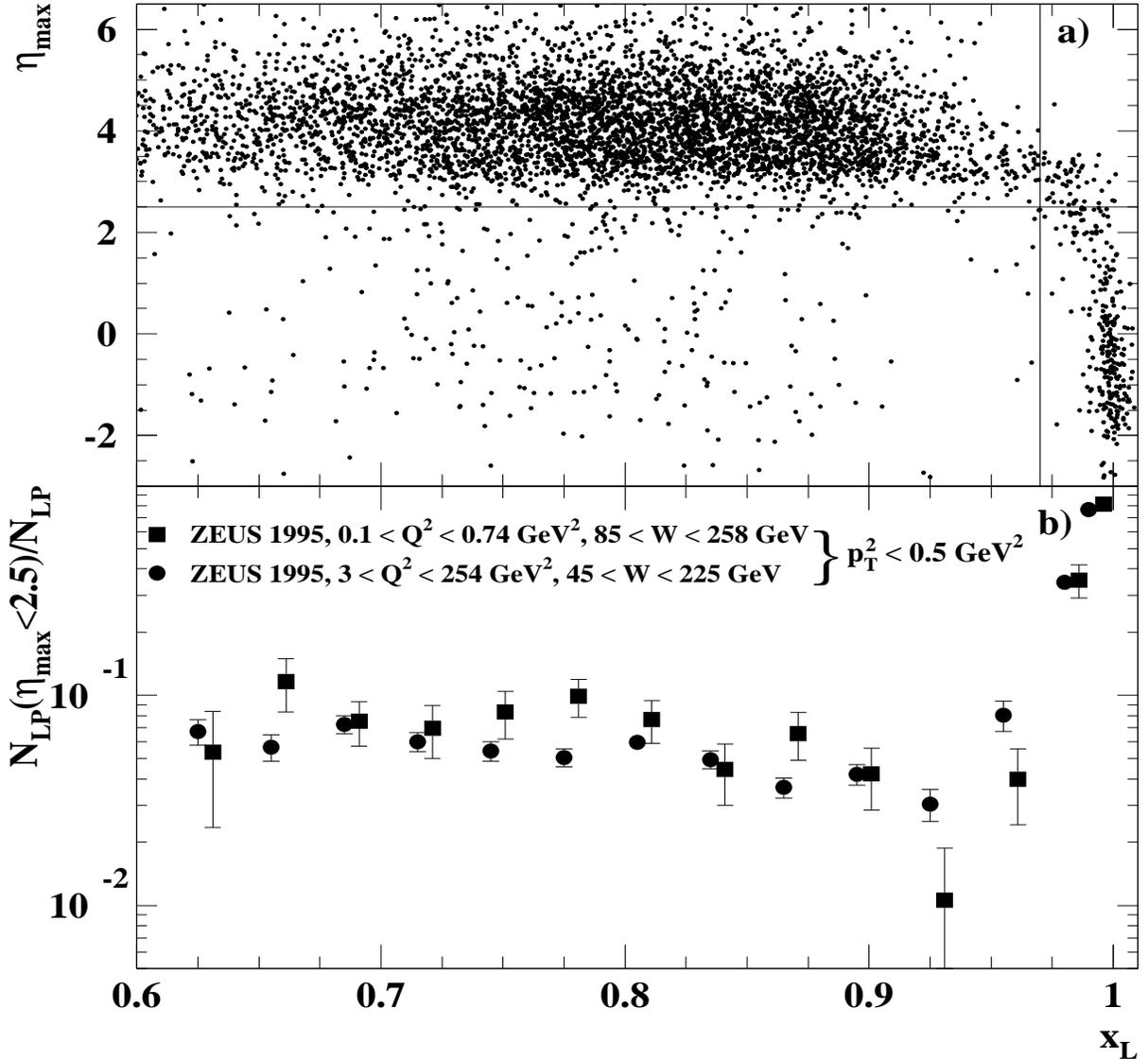}
}
\end{center}
\vspace{-0.1cm}
\caption{a) Distribution of the DIS events in the
($\eta_{\max}$, $x_L$) plane. The horizontal and vertical lines indicate    
$\eta_{\max}=2.5$ and $x_L=0.97$, respectively. b) Fraction of events with 
$\eta_{\max}<2.5$ for both the BPC and the DIS samples. 
The BPC data are slightly shifted in $x_L$ for clarity of 
presentation. The bars 
indicate the statistical uncertainties; systematic uncertainties mostly 
cancel in the ratio. }
\label{fig:etamax}
\end{figure}

Figure~\ref{fig:etamax}a) shows the distribution of the DIS events in the 
($\eta_{\max}$, $x_L$) plane, where $\eta_{\max}$ is the pseudorapidity of the 
most-forward energy deposit of at least 400~MeV in the CAL. 
The accumulation of events at $x_L \approx 1$, which mostly
have $\eta_{\max} < 2.5$, 
is due to diffractive events~\cite{earlydiffraction},
$e^+p \rightarrow e^+Xp$, in which 
the final-state proton remains intact and carries
approximately the same momentum as the incoming proton. 
Events with $\eta_{\max}<2.5$ and $x_L~\ltap~0.97$ are
ascribed to double diffractive 
dissociation, $e^+p \rightarrow e^+XN$, where the proton  
dissociates into the state $N$, with mass $M_N$. 
Although $N$ is produced at $x_L \simeq 1$, 
the proton from the decay of $N$ has a lower value of $x_L$.
When  both $M_N$ and $M_X$ are small, the systems $X$
and $N$ are separated by a large gap in pseudorapidity. 
If $M_N< 4$-5~GeV, only the proton from the system $N$ is measured, while the
other particles escape undetected down the beam pipe. The topology of a  
doubly dissociative event is thus characterised by a rapidity gap in the 
forward region in conjunction with a low-$x_L$ proton.

To select a sample of diffractive events, the requirement 
$\eta_{\max}<2.5$ was imposed. Figure~\ref{fig:etamax}b) shows the
fraction of BPC and DIS events with a leading proton that also have
$\eta_{\max}<2.5$ (see also Table~\ref{tab-fig5}). Events with a large
rapidity gap 
dominate for $x_L \approx 1$. For $0.6 < x_L <0.97$, the
fraction of leading proton events with a large rapidity gap is less
than 10\% in any given bin, and is 
only weakly dependent on $Q^2$ and $x_L$. This result, which indicates 
that diffraction is not the main mechanism responsible for
 production of leading protons in this region, is consistent with the
predictions of the Regge-based models~\cite{SNS, SNS1}.

\subsection{Momentum spectra of leading protons}

\subsubsection{Longitudinal-momentum spectra}
\label{xlspectra}

\begin{figure}[htb]
\begin{center}
\leavevmode
\hbox{%
\epsfxsize = 16cm
\epsfysize = 16cm
\epsffile{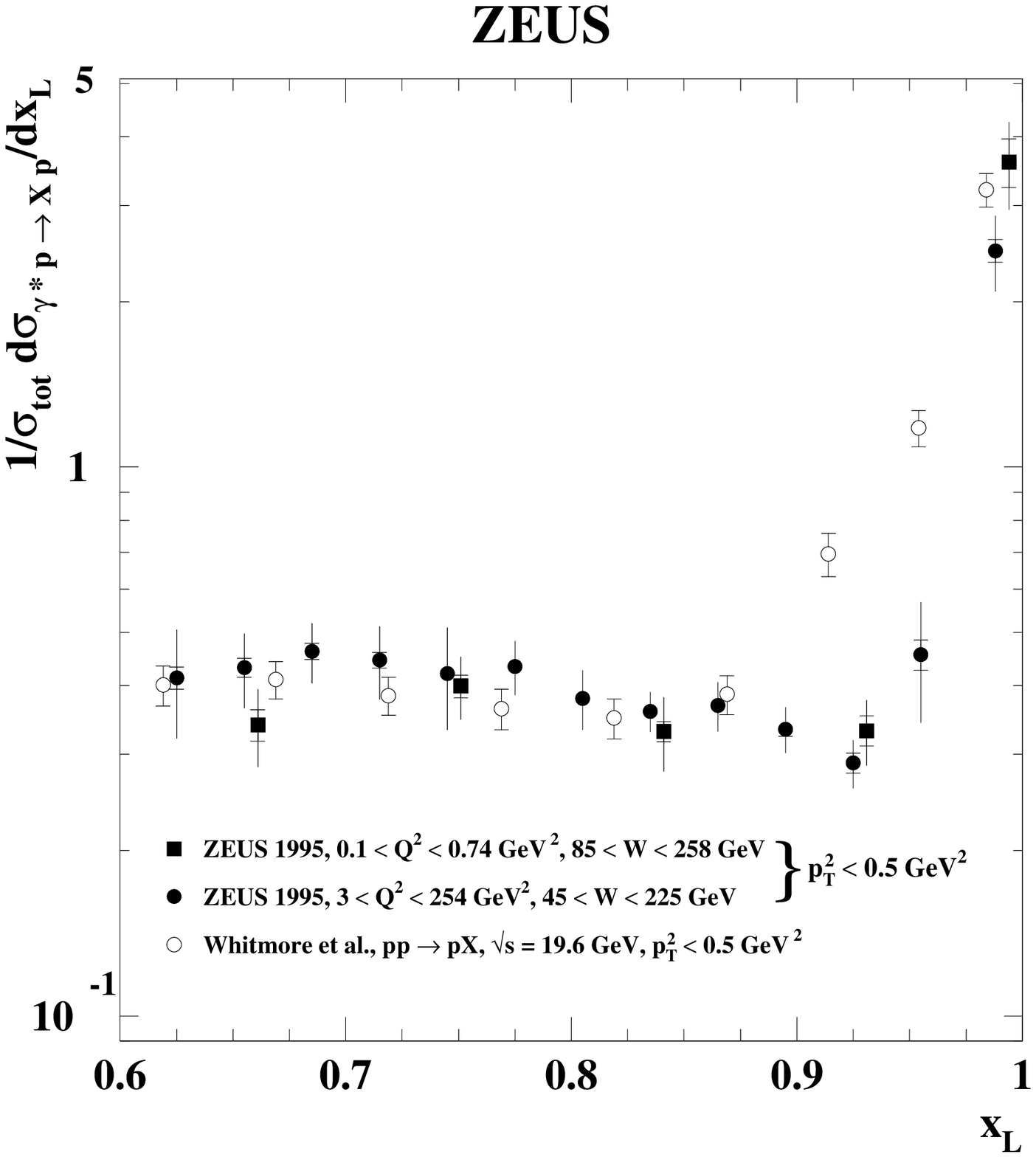}
}
\end{center}
\caption{{\protect{The normalised cross-section $(1/\sigma_{\rm tot}) \cdot
d\sigma_{\gamma^* p \rightarrow Xp}/dx_L$ for the BPC and DIS data
compared to the $pp$ data~\protect\cite{whitmore} in the
region $p_T^2<0.5~\mathrm{GeV}^2$. 
The inner bars indicate the statistical uncertainties and the outer bars
are the statistical and systematic uncertainties summed in quadrature.}}}
\label{dsigmadxl_fermilab}
\end{figure}

The normalised cross-section $r^{\rm LP(3)}=(1/\sigma_{\rm tot}) \cdot
d\sigma_{\gamma^*p \rightarrow Xp}/dx_L$ for the reaction 
$e^+p \rightarrow e^+Xp$ with a leading proton having $x_L>0.6$ and 
$p_T^2 <  0.5$~GeV$^2$ is shown in Fig.~\ref{dsigmadxl_fermilab} and
given in Table~\ref{tab-fig6} for the BPC sample, integrated over the 
range $0.1 <Q^2< 0.74$~GeV$^2$, $85<W<258$~GeV, $1.5 \times 10^{-6} < x < 1.0 
\times 10^{-4}$, and for the DIS sample, integrated over the region 
$3<Q^2< 254$~GeV$^2$, $45<W<225$~GeV, $1.2 \times 10^{-4} < x < 4 \times 10^{-2}$.
These results are compared with those from the reaction 
$pp \rightarrow pX$ 
at $\sqrt{s}= 19.6$~GeV~\cite{whitmore}, 
integrated over the same $p_T^2$ region and normalised to the
corresponding inelastic cross section. For $x_L~\ltap$~0.9, the 
fraction of events with a leading proton is consistent for
the $pp$ and $\gamma^{*}p$ data sets, in accord with vertex
factorisation.
A dependence on the centre-of-mass energy  is apparent for
$x_L>0.9$,
as expected from Regge parametrisations of the $pp \rightarrow pX$ 
data~\cite{ganguli,batista}.

\begin{figure}[htb]
\begin{center}
\leavevmode
\hbox{%
\epsfxsize = 16cm
\epsfysize = 16cm
\epsffile{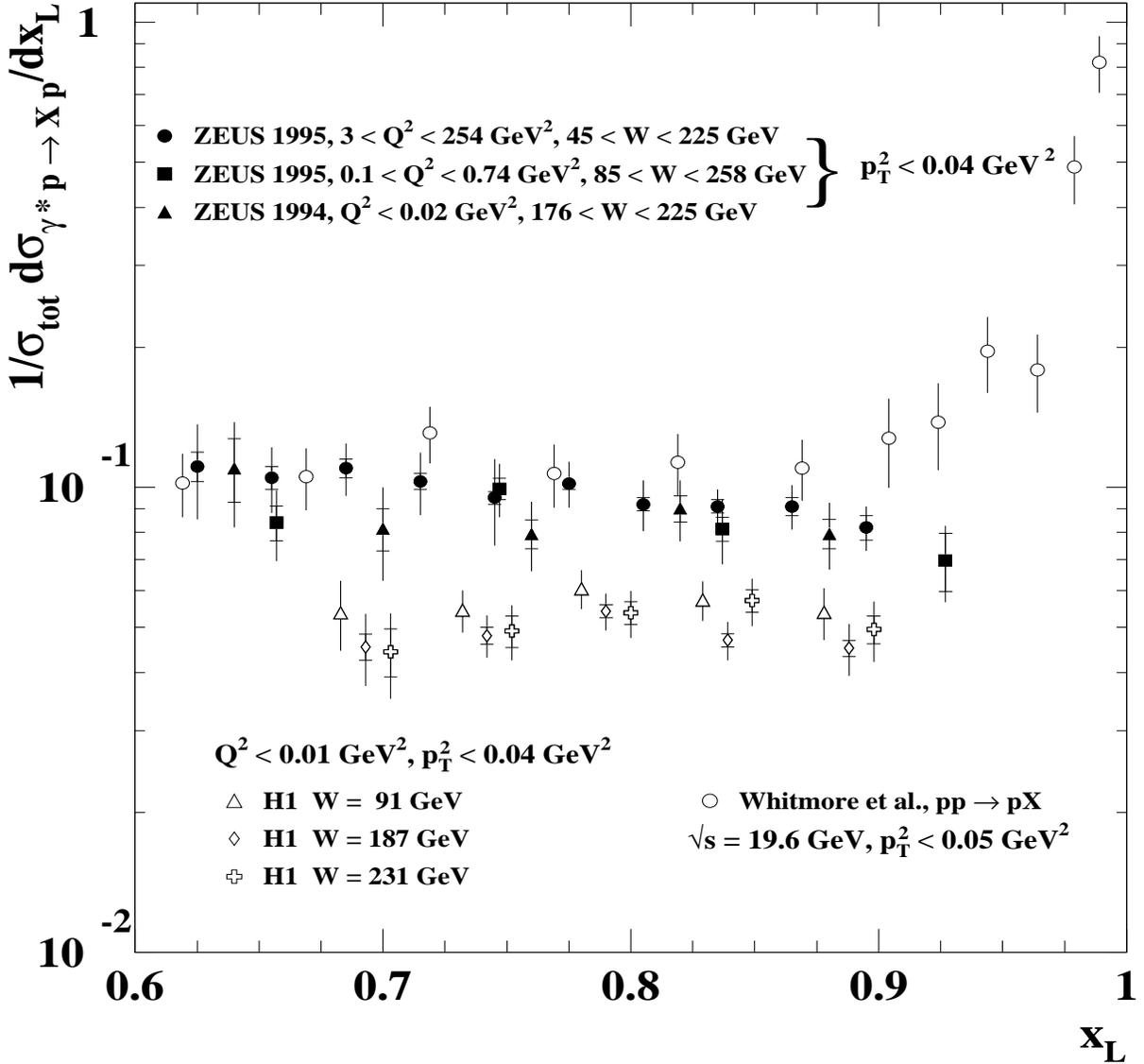}
}
\end{center}
\caption{{\protect{
The normalised cross-section 
$(1/\sigma_{\rm tot}) \cdot d\sigma_{\gamma^* p \rightarrow Xp}/dx_L$
for the photoproduction, BPC and DIS data  compared to 
the $pp$ data~\protect\cite{whitmore} in the region 
$p_T^2<0.04~\mathrm{GeV}^2$. 
The inner bars indicate the statistical 
uncertainties and the outer bars are the statistical and systematic uncertainties 
summed in quadrature. 
The H1 results~\protect\cite{h1php} are also shown.
}}}
\label{dsigmadxl_php}
\end{figure}

Figure~\ref{dsigmadxl_php} and Table~\ref{tab-fig7} present the
photoproduction, BPC and DIS data
for the lower $p_T^2$ region, $p_T^2<0.04$~GeV$^2$ and $0.6 < x_L < 0.95$, 
where the upper cut on $x_L$, which removes the diffractive events, is
set by the LPS acceptance for this $p_T^2$ range.
The fraction of events with a leading proton is approximately the same in
all three regimes.
The $pp$ data~\cite{whitmore} for $p_T^2<0.05$ GeV$^2$ 
again agree with the $ep$ data for
$x_L$~$\ltap$~0.9.
The present photoproduction results, however, are significantly higher
than those found by H1~\cite{h1php} in
similar ranges of $Q^2$, $W$ and $p_T^2$.

\begin{figure}[htb]
\begin{center}
\leavevmode
\hbox{%
\epsfxsize = 16cm
\epsfysize = 16cm
\epsffile{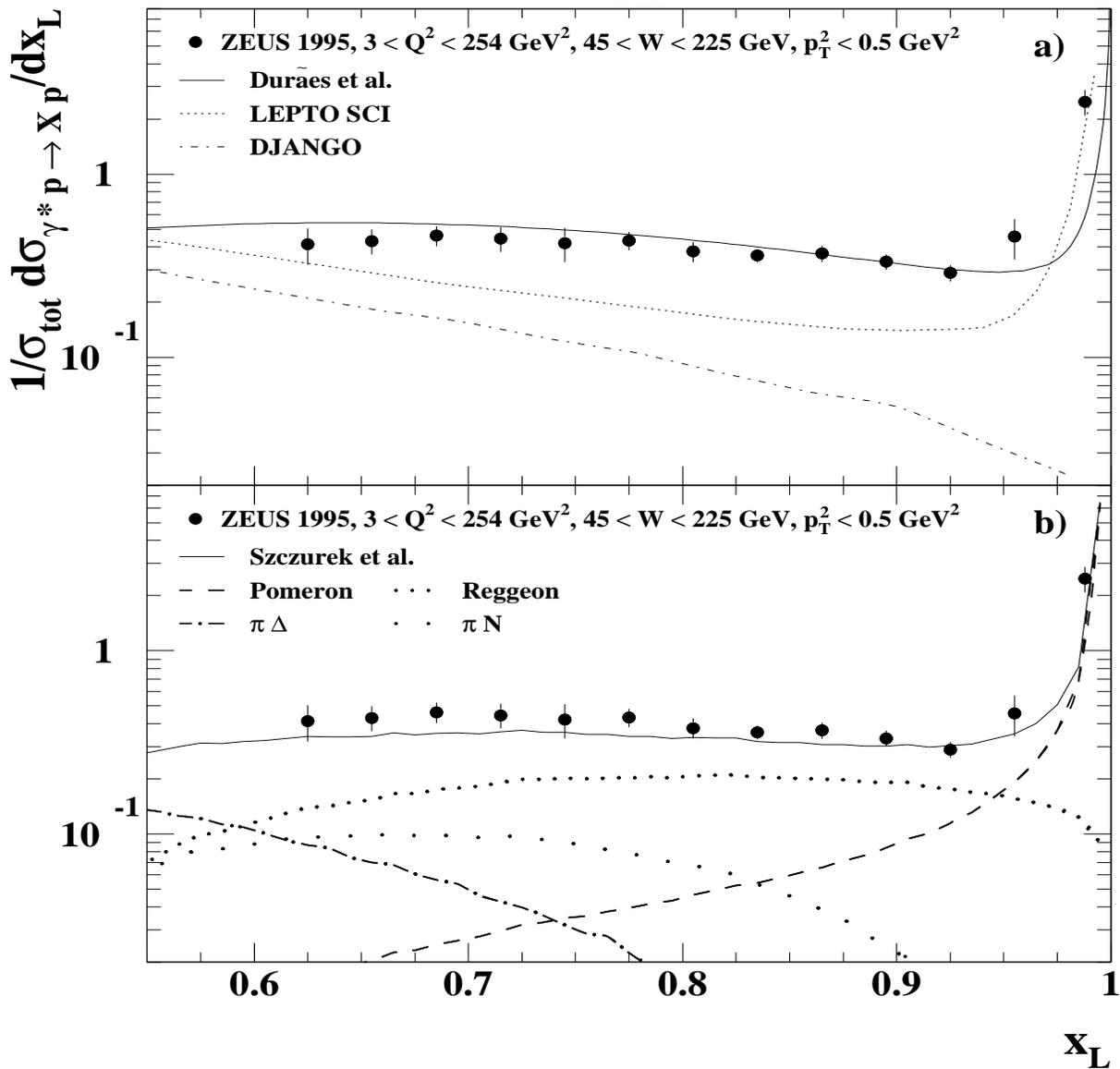}
}
\end{center}
\caption{{\protect{The normalised cross-section 
$(1/\sigma_{\rm tot}) \cdot d\sigma_{\gamma^* p \rightarrow Xp}/dx_L$ 
for the DIS data (as shown in Fig.~\protect\ref{dsigmadxl_fermilab}) 
compared to a) the model of Dur\~aes et al. (solid curve), LEPTO6.5 
(dashed curve) and DJANGO (dot-dashed curve), and b)
to the model of Szczurek  et al.  (solid curve). For the latter, 
the individual contributions of Pomeron, Reggeon and pion exchanges are 
indicated; for pion exchange, the contribution of final states with 
isospin $I=1/2$ and $I=3/2$ are shown separately. 
}}}
\label{dsigmadxl_models}
\end{figure}

Figure~\ref{dsigmadxl_models} compares the DIS data to the specific models 
of Section~\ref{models}.
The standard DIS Monte Carlo generator DJANGO~\cite{django}
predicts a stronger decrease of the cross section with $x_L$ 
than that observed and, in addition,
has no diffractive peak. It also substantially underestimates the
rate of leading proton production in the measured $x_L$ range.
The SCI model~\cite{sci}, as
implemented in LEPTO6.5 \cite{lepto}, also falls below the data, even 
though it generates a larger number of leading protons than DJANGO
and has a peak at $x_L= 1$. These two models are
ruled out by the data.

The QCD-inspired model of Dur\~aes et al.~\cite{igm,igm_pc}
is in better agreement with the data, but is too low in the diffractive 
peak region. Nevertheless, the similarity to
the data is remarkable, given the small number of free parameters in the
model.

The Regge-based calculation of Szczurek et al.~\cite{SNS} 
agrees with the data, although it is  somewhat 
too low at small values of $x_L$. In this approach, 
leading proton production for $0.6 < x_L < 0.9$ is dominated by 
isoscalar Reggeon exchange; diffractive processes, due to Pomeron
exchange, 
become increasingly important as $x_L$ approaches unity.
The contribution of pion exchange, including the production of $\Delta$
baryons, is also evaluated.
The normalisation of the Reggeon contribution has a 
large theoretical uncertainty~\cite{SNS,kolya_private} and 
the present data suggest that this contribution  
should be increased with respect to the assumption made 
by Szczurek et al.~\cite{SNS}. 

\subsubsection{Transverse-momentum spectra}
\label{ptspectra}

\begin{figure}[htbp!]
\epsfysize=16cm
\epsfysize=16cm
\epsffile{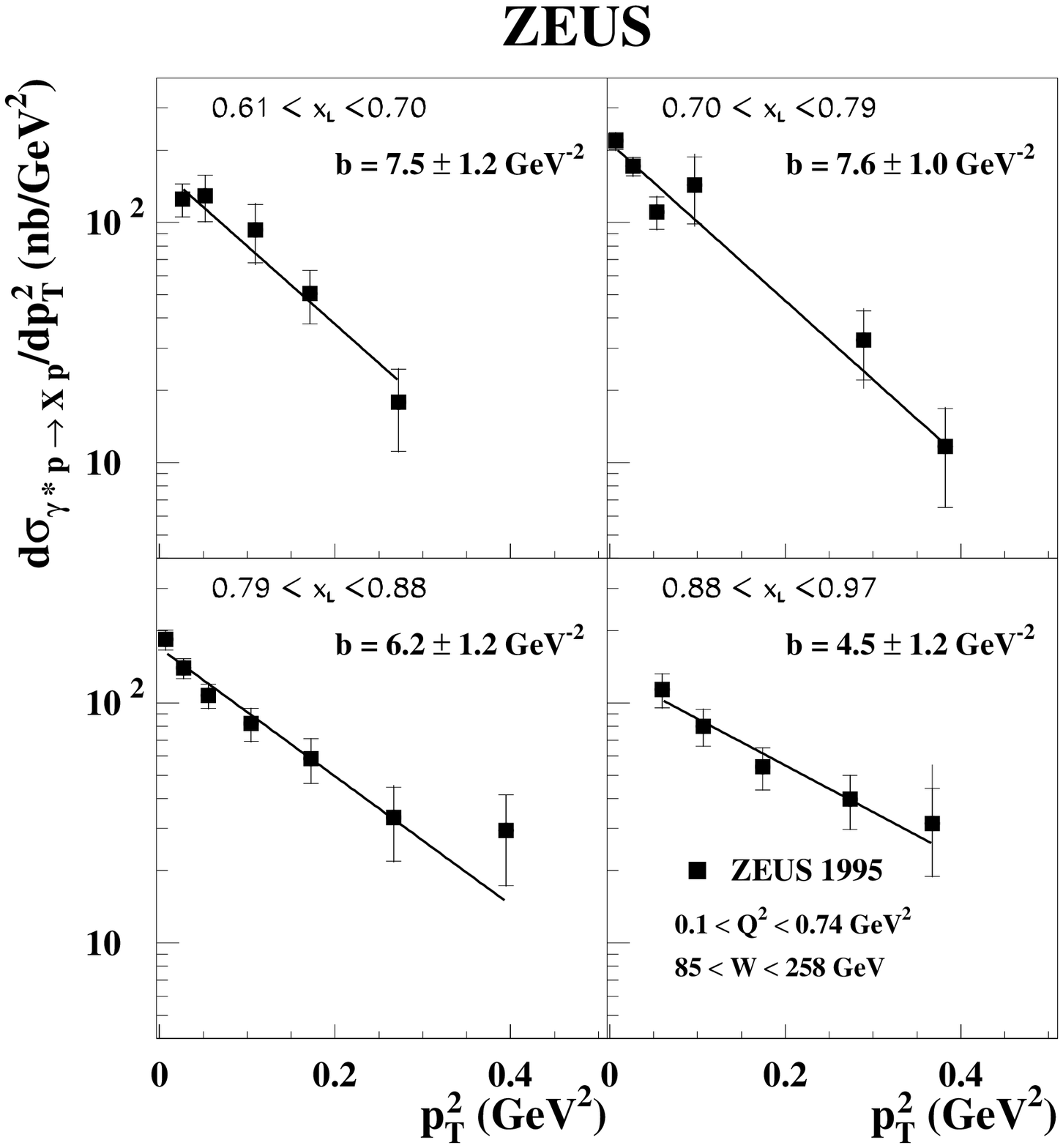}
\caption{
The differential cross-section $d\sigma_{\gamma^* p \rightarrow Xp}/dp_T^2$ 
for several $x_L$ bins for the BPC sample. 
The lines represent the results of fits to the functional form
$d\sigma_{\gamma^* p \rightarrow Xp}/dp_T^2 \propto e^{-b p_T^2}$. 
The fitted values of $b$ and their statistical uncertainties are also
given. The inner  bars indicate the size of the statistical uncertainties,
the outer bars show the statistical and systematic uncertainties summed in 
quadrature.
}
\label{lpspt_bpc}
\end{figure}

\begin{figure}[htbp!]
\epsfxsize=16cm
\epsfysize=16cm
\epsffile{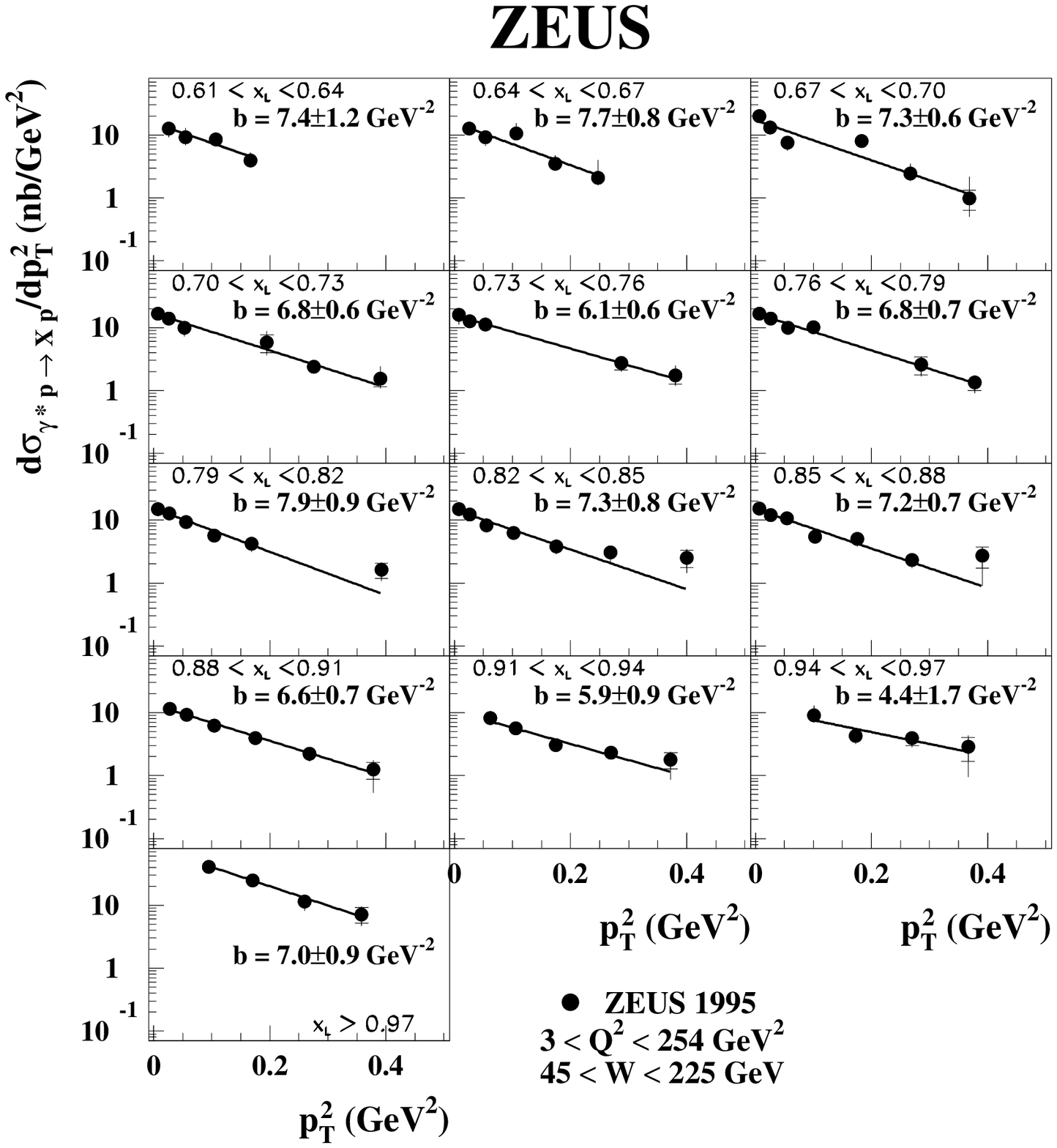}
\caption{
The differential cross-section $d\sigma_{\gamma^* p \rightarrow Xp}/dp_T^2$
 for several $x_L$ bins for the DIS sample. 
The lines represent the results of fits to the functional form
$d\sigma_{\gamma^* p \rightarrow Xp}/dp_T^2 \propto e^{-b p_T^2}$. 
The fitted values of $b$ and their statistical uncertainties are also given. 
The inner  bars
indicate the size of the statistical uncertainties, the outer bars show the 
statistical and systematic uncertainties summed in quadrature.
}
\label{lpspt}
\end{figure}

\begin{figure}[htbp!]
\epsfxsize=16cm
\epsfysize=16cm
\centerline{\epsffile{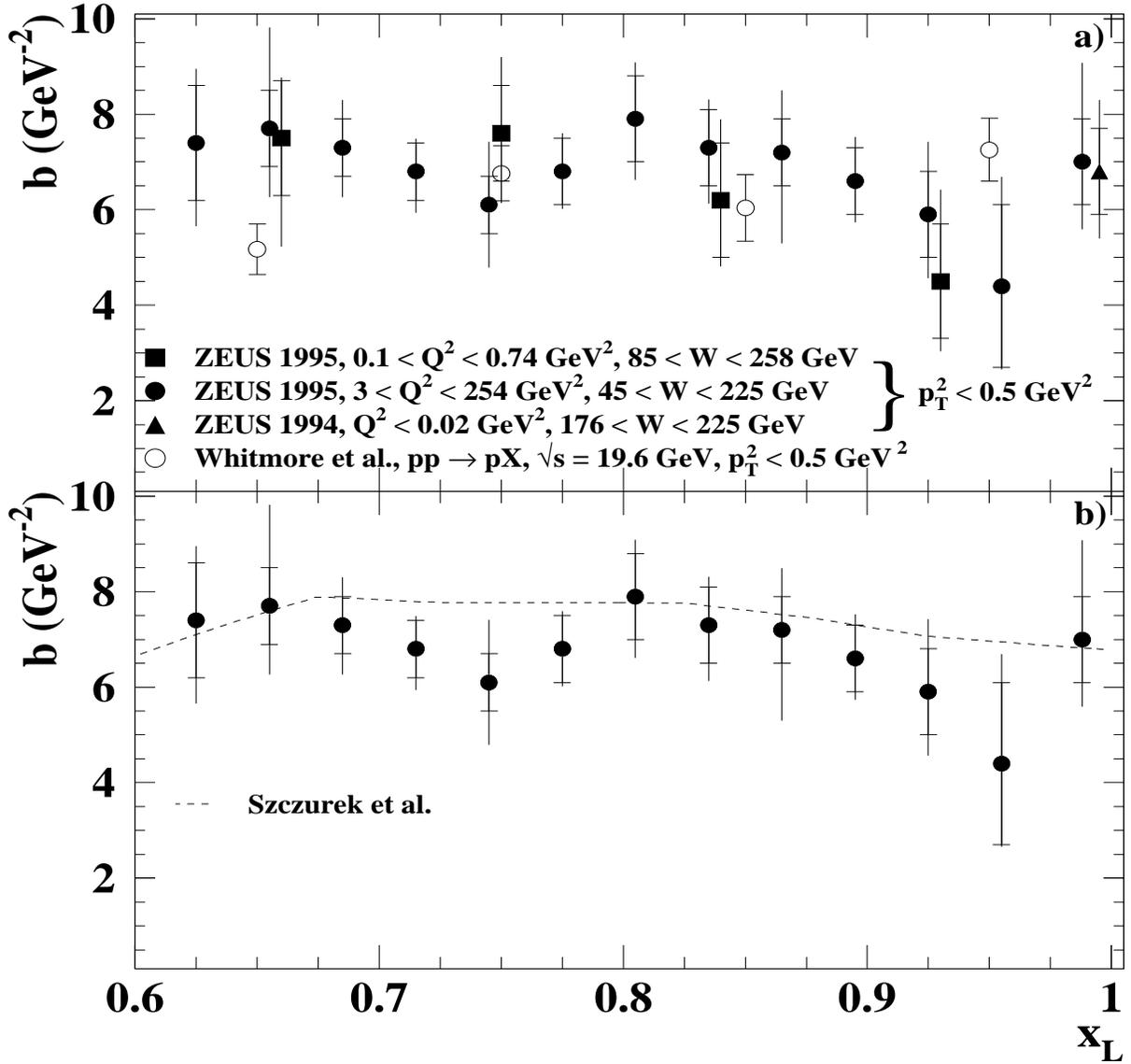}}
\vspace{-0.1cm}
\caption{a) The slopes, $b$, of the $p_T^2$ distributions
for leading protons as a function of $x_L$ for the BPC and DIS data samples.
For clarity of presentation, the BPC points are plotted slightly 
shifted in $x_L$. The inner bars indicate the
statistical uncertainties and the outer are the statistical and systematic 
uncertainties summed in quadrature. The photoproduction result at
$x_L \simeq 1$ is also shown, as are the data from the reaction $pp
\rightarrow pX$ at 
$\sqrt{s}=19.6~\mathrm{GeV}$. b) The slopes, $b$, of the $p_T^2$ 
distributions
for leading protons as a function of $x_L$ for the DIS data sample, 
compared with the prediction of Szczurek et al. (dashed line).
}
\label{lpsb}
\end{figure}

The cross-sections $d\sigma_{\gamma^*p \rightarrow Xp}/dp_T^2$ 
are shown in Fig.~\ref{lpspt_bpc} for
the BPC sample integrated over the range $0.1 <Q^2< 0.74$~GeV$^2$,
$85<W<258$~GeV, $1.5 \times 10^{-6} < x < 1.0 \times 10^{-4}$
for different $x_L$ selections.
Similar distributions for the DIS sample are shown in Fig.~\ref{lpspt},
integrated over the region $3<Q^2< 254$~GeV$^2$, $45<W<225$~GeV,
$1.2 \times 10^{-4} < x < 4 \times 10^{-2}$.
In all regions, the form 
$d\sigma_{\gamma^*p \rightarrow Xp}/dp_T^2 = A e^{-bp_T^2}$ fits 
satisfactorily the data.
The values of the slope-parameters $b$ obtained for each $x_L$ bin 
are, within uncertainties, independent of $x_L$, as shown in Fig.~\ref{lpsb}
 (see also Table~\ref{tab-fig11}).

The BPC and DIS data together indicate that $b$ is independent of 
$Q^2$ and $x_L$.
The mean value of $b$ for the BPC data is 
$\langle b \rangle=6.6 \pm 0.6~{\mbox (\rm stat.)} \pm 0.8~{\mbox 
(\rm syst.)}$~GeV$^{-2}$ and 
$\langle b \rangle=6.9 \pm 0.2~{\mbox (\rm stat.)} \pm 0.8~{\mbox (\rm syst.)} 
$~GeV$^{-2}$ 
for the DIS data for $0.6 < x_L < 0.97$.
Also plotted in Fig.~\ref{lpsb}a) is the result obtained for diffractive 
photoproduction~\cite{lps94}, which is consistent with the values found 
at higher $Q^2$.
In addition, the present results are compatible with the $pp
\rightarrow pX$ data~\cite{whitmore},
also shown in Fig.~\ref{lpsb}a). This, together with 
the fact that $b$ is approximately $Q^2$-independent, provides
additional support for vertex factorisation.

The predictions of Szczurek et al.~\cite{SNS} are in accord with the
transverse-momentum data, as shown in Fig.~\ref{lpsb}b).
In this model, no
$Q^2$ dependence is expected for $b$~\cite{kolya_APP}.
The DJANGO program (not shown) has an effective slope
$b \approx 4$~GeV$^{-2}$. 
LEPTO6.5 (also not shown) has $b \approx 3.5$~GeV$^{-2}$.
Both are independent of $x_L$ and significantly below the values measured. 
The model of Dur\~aes et al.~\cite{igm} does not make explicit predictions 
for the transverse-momentum distribution.

\subsection{The proton-tagged structure-function
\boldmath{$\bar{F}_2^{\rm LP}$}}
\label{f2}

The acceptance-corrected fraction of events with a leading proton 
is used to measure the proton-tagged
structure-function $\bar{F}_2^{\rm LP}$, as discussed in
Section~\ref{ratio-method}. To select a predominantly non-diffractive sample, the cut $x_L<0.97$ was
imposed.

\begin{figure}[htbp!]
\centerline{\epsffile{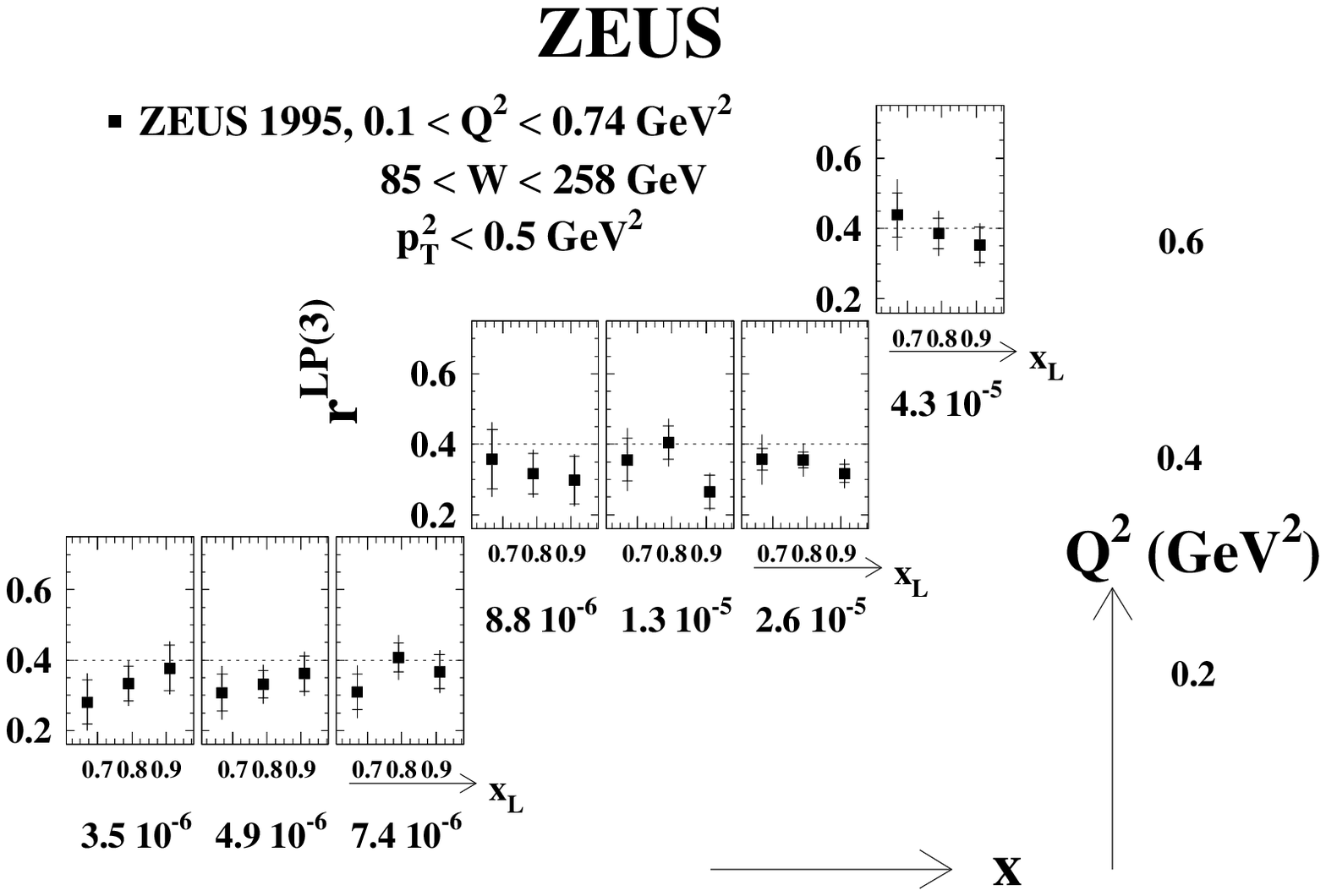}}
\vspace{-0.1cm}
\caption{  The ratio $r^{\rm LP(3)}=\bar{F}_2^{\rm LP(3)}/F_2$ as a
function of
  $x_L$ in bins of $x$ and $Q^2$ (BPC sample), for protons with 
  $p_T^2 < 0.5~\mathrm{GeV}^2$.  The inner 
  bars show the statistical
  uncertainties and the outer bars the statistical and systematic 
uncertainties
  added
  in quadrature. The dashed line $r^{\rm LP(3)}=0.4$ is overlaid to guide
the eye. 
}
\label{ratio_xl_bpc} \end{figure}

\begin{figure}[htbp!]
\centerline{\epsffile{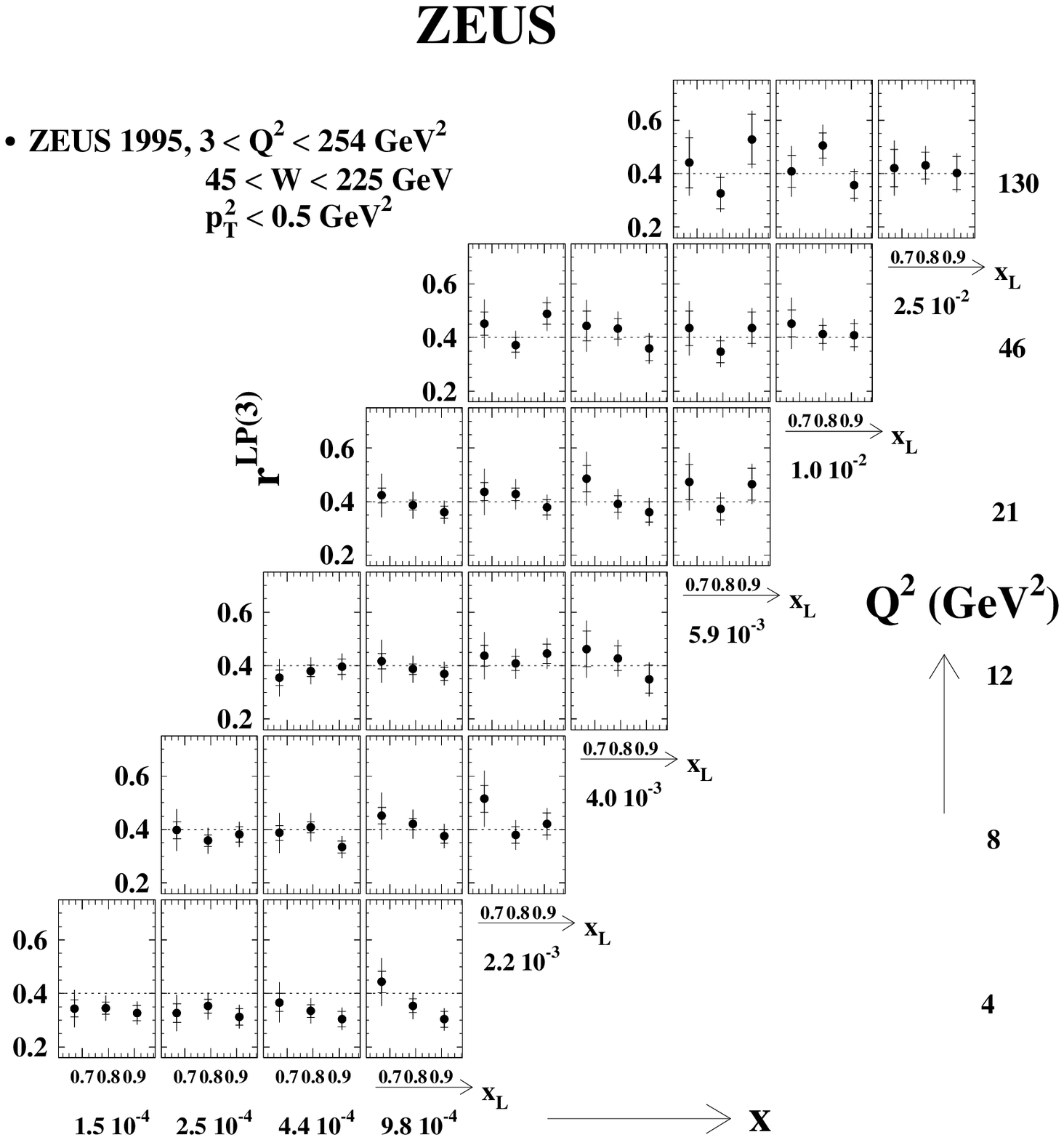}}
\vspace{-0.1cm}
\caption{  The ratio $r^{\rm LP(3)}=\bar{F}_2^{\rm LP(3)}/F_2$ as a function of
  $x_L$ in bins of $x$ and $Q^2$ (DIS sample), for protons with 
  $p_T^2 < 0.5~\mathrm{GeV}^2$.  The inner  
  bars show the statistical
  uncertainties and the outer bars the statistical and systematic 
uncertainties 
  added in quadrature. The dashed line $r^{\rm LP(3)}=0.4$ is overlaid to
guide the eye. }
\label{ratio_xl_dis}
\end{figure}

\begin{figure}[htbp!]
\centerline{\epsffile{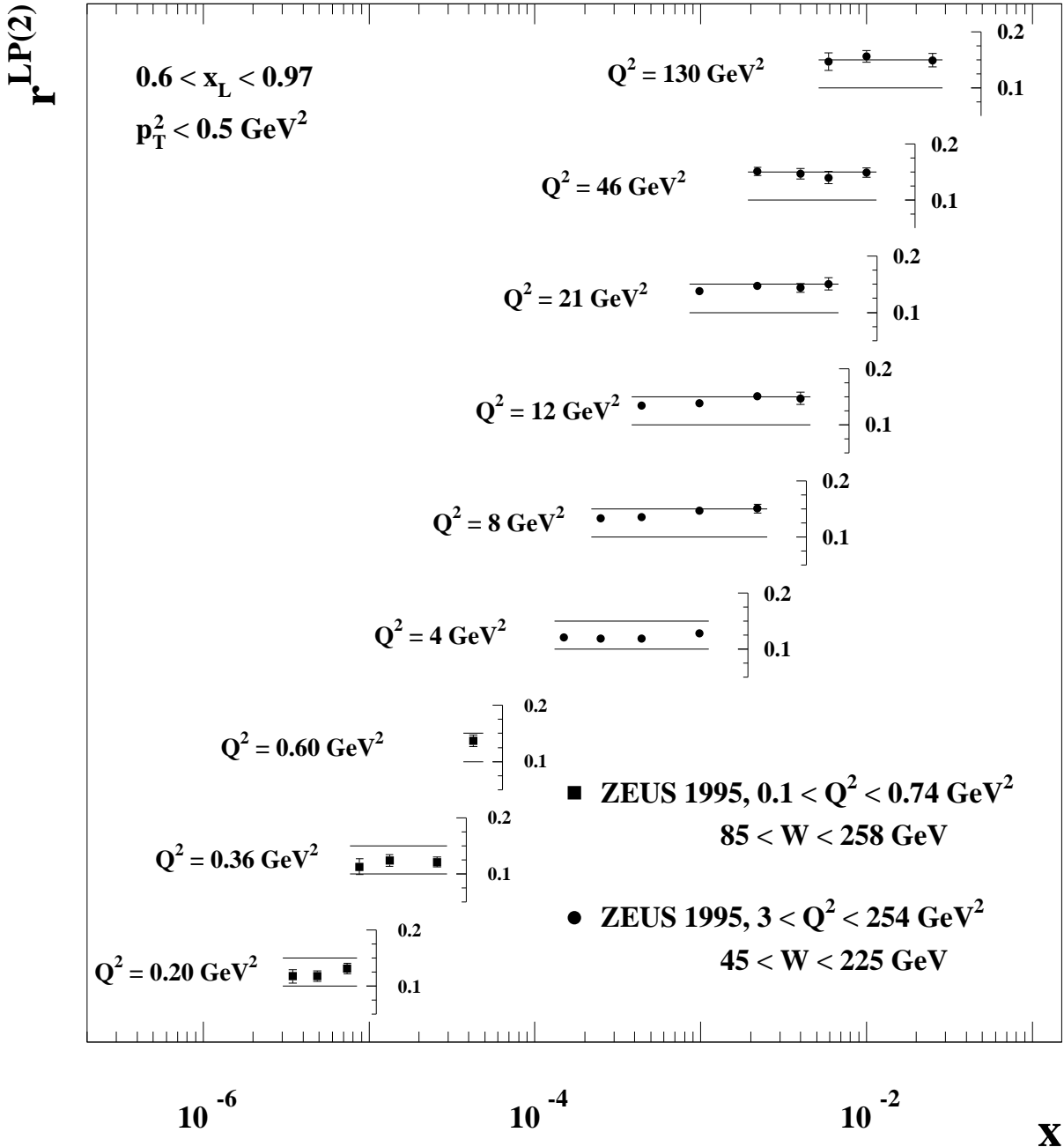}}
\vspace{-0.1cm}
\caption{The ratio $r^{\rm LP(2)}=\bar{F}_2^{\rm LP(2)}/F_2$
  as a function of $x$ for fixed $Q^2$ values, for protons with $0.6 < x_L <
  0.97$ and $p_T^2 < 0.5~\mathrm{GeV}^2$.  
The error 
  bars show
  the statistical uncertainties. A fully correlated systematic uncertainty
of $\pm 13\%$ is not shown.
The horizontal lines 
  $r^{\rm LP(2)}=0.10$ and $r^{\rm LP(2)}=0.15$ are overlaid to guide the eye.
}
\label{lps_yx}
\end{figure}

Figures~\ref{ratio_xl_bpc} and~\ref{ratio_xl_dis} show 
$r^{\rm LP(3)}(x,Q^2, x_L)$, determined using Eq.~(\ref{eq:rlp3}),
for the BPC and DIS samples, respectively, in several $x$ and $Q^2$ bins
for $p_T^2<0.5$~GeV$^2$. The data are also given in
Tables~\ref{tab-fig12}-\ref{tab-fig13b}.
Only a weak dependence on $x$ and $Q^2$ is apparent, indicating that 
$\bar{F}_2^{\rm LP(3)}$ has approximately the same $x$ 
and $Q^2$ dependence as $F_2(x,Q^2)$.
The data exhibit a weak $x_L$ dependence, as already seen in
Fig.~\ref{dsigmadxl_fermilab}.
Figure~\ref{lps_yx} shows $r^{\rm LP(2)}$ for fixed $Q^2$
values as a function of $x$; the results again have little 
$x$ dependence (see also Table~\ref{tab-fig14}).

\begin{figure}[htbp!]
\epsfxsize=16cm
\epsfxsize=16cm
\centerline{\epsffile{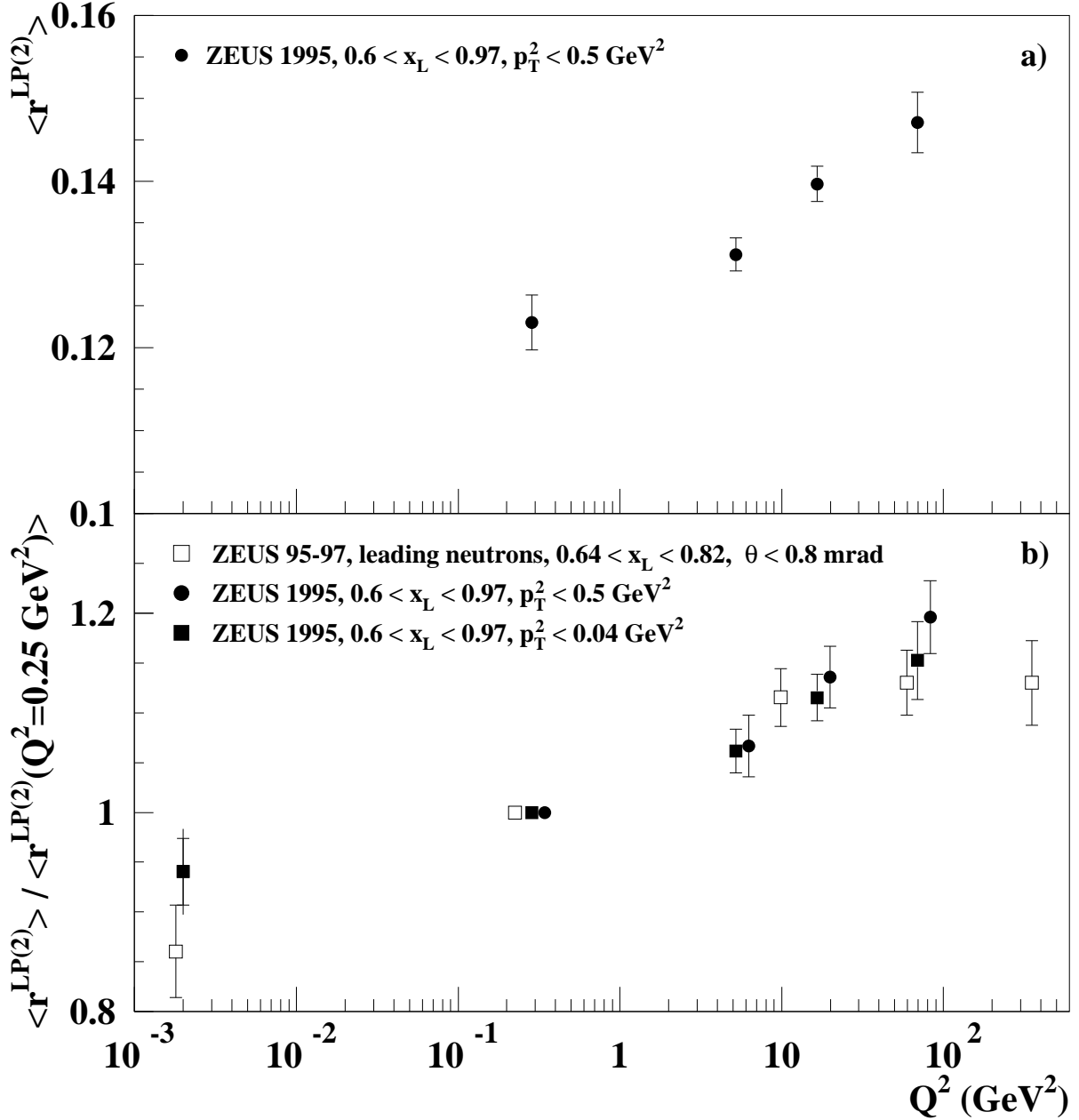}}
\vspace{-0.1cm}
\caption{The average ratio 
$\langle{r}^{\rm LP(2)}\rangle=\bar{F}_2^{\rm LP(2)}/F_2$ as a function of 
$Q^2$. The error bars show the statistical uncertainties. 
A fully correlated systematic uncertainty of 13\% is not shown.
a) $\langle{r}^{\rm LP(2)}\rangle$ for the range  
$0.6<x_L<0.97$ and $p_T^2 < 0.5~\mathrm{GeV}^2$.
b) $\langle{r}^{\rm LP(2)}\rangle$ as
a function of $Q^2$ for two different $p_T^2$ ranges normalised to the
value at $Q^2=0.25~\mathrm{GeV}^2$. The error bars show the statistical
uncertainties; systematic errors mostly cancel in the ratio. The ZEUS data 
for  leading neutron production, also normalised to the value at
$Q^2=0.25~\mathrm{GeV}^2$, are also shown. The points for 
$p_T^2<0.5~\mathrm{GeV}^2$
are slightly shifted for clarity of presentation.}
\label{q2dep}
\end{figure}

Figure~\ref{q2dep}a) and Table~\ref{tab-fig15a} present the BPC and DIS
data  
for $0.6<x_L<0.97$ and $p_T^2<0.5$ GeV$^2$, averaged over $x$
for different $Q^2$ ranges, $\langle {r}^{\rm LP(2)}(Q^2) \rangle$.
The leading proton yield increases by approximately 20\%, 
from $\langle {r}^{\rm LP(2)} \rangle \approx 0.12$ to  $\langle {r}^{\rm LP(2)}
\rangle \approx 0.15$, when $Q^2$ varies from $\approx
0.25$~GeV$^2$ (the average value of $Q^2$ for the BPC sample) to
100~GeV$^2$, 
indicating a modest but definite breakdown of vertex factorisation. 
Figure~\ref{q2dep}b) and Table~\ref{tab-fig15b} present the points of
Fig.\ref{q2dep}a) normalised to 
$\langle {r}^{\rm LP(2)}(Q^2=0.25$~GeV$^2) \rangle$. The results for the
restricted $p_T^2$ range, $p_T^2<0.04$~GeV$^2$, are also shown. The 
breaking of vertex factorisation
is approximately the same for $p_T^2<0.5$~GeV$^2$ and
for $p_T^2<0.04$~GeV$^2$. An effect of similar size was measured for 
leading neutron production~\cite{neutrons}; the corresponding data,
normalised to the value at $Q^2 = 0.25$ GeV$^2$, are
also shown in Fig.~\ref{q2dep}b).
The neutron data are measured for scattering angles less than 0.8 mrad,
corresponding to $p_T^2 < 0.43 \cdot x_L^2$~GeV$^2$.

This $Q^2$ dependence of the proton yield 
can be qualitatively ascribed to absorptive effects 
in the $\gamma^{\ast}p$ system~\cite{dalesio}. The transverse size of the 
virtual photon decreases 
with increasing $Q^2$, reducing the likelihood that the produced baryon 
rescatters on the hadronic component of the virtual photon. 

\begin{figure}[htbp!]
\epsfxsize=16cm
\epsfysize=16cm
\centerline{\epsffile{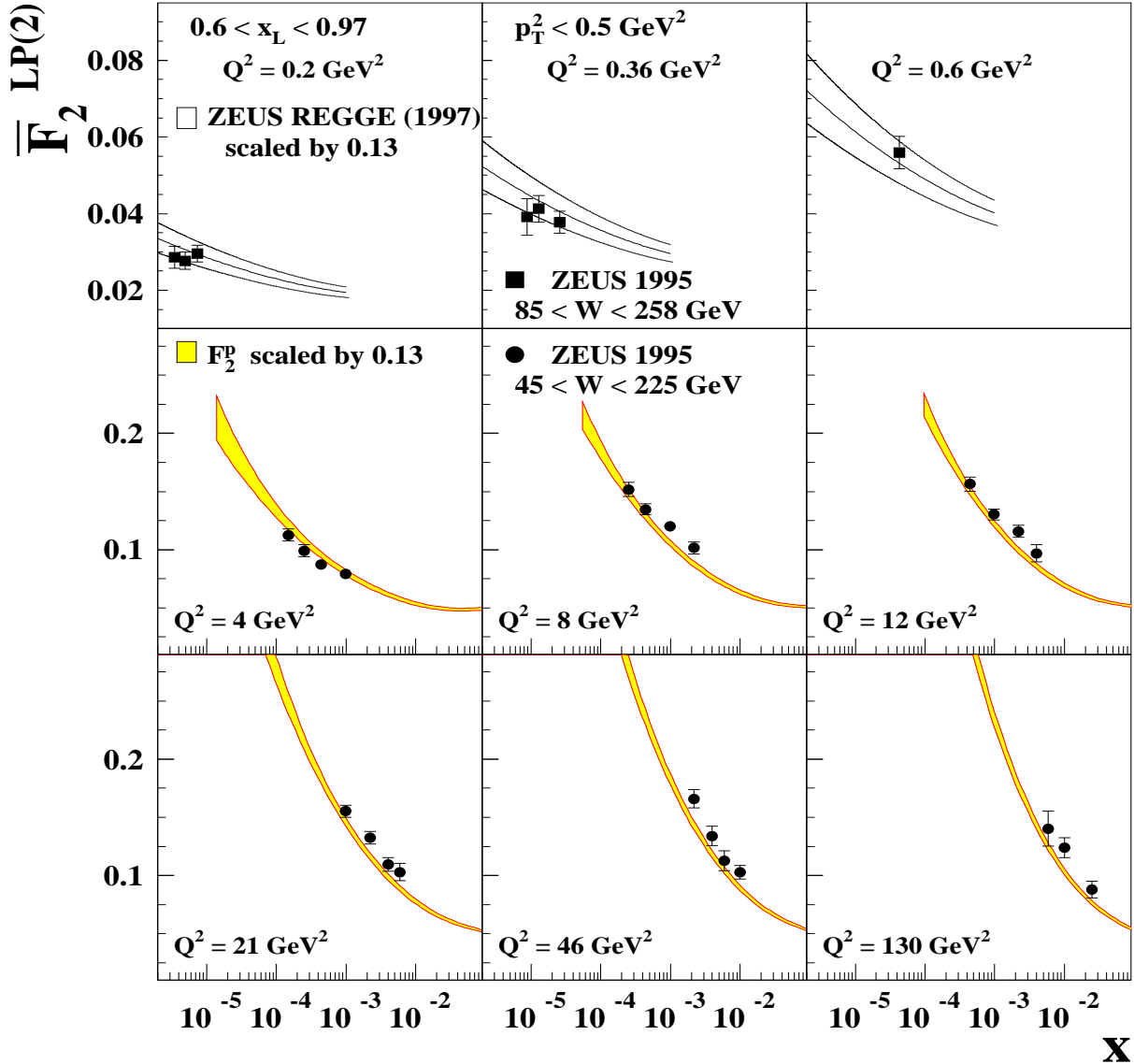}}
\vspace{-0.1cm}
\caption{
The structure-function $\bar{F}_2^{\rm LP(2)}$ as a function of $x$ 
for
$0.6<x_L<0.97$ and $p_T^2<0.5~\mathrm{GeV}^2$. The  bands show the
one-standard-deviation limits of the $F_2$ parametrisations 
used, scaled by the
average value of $r^{\rm LP(2)}$ ($\langle r^{\rm LP(2)} \rangle \simeq 0.13$). 
The error 
  bars show
  the statistical uncertainties. A fully correlated systematic uncertainty
of $\pm 13\%$ is not shown.}
\label{absolute_f2} 
\end{figure}

The data of Fig.~\ref{lps_yx} are presented in Fig.~\ref{absolute_f2} in 
terms of  $\bar{F}_2^{\rm LP(2)}$,
obtained by multiplying $r^{\rm LP(2)}$ by  $F_2$.
For the BPC region, a parameterisation of the ZEUS
$F_2$ results~\cite{zeusbpt} was used. For the DIS region, the 
parameterisation of Botje~\cite{michiel} was used. Since  
$r^{\rm LP(2)}$ is approximately independent of $Q^2$ and $x$, 
$\bar{F}_2^{\rm LP(2)}$ is approximately 
proportional to $F_2$. As indicated in the figure, the proportionality 
constant between $\bar{F}_2^{\rm LP(2)}$ and $F_2$ is 
$\langle r^{\rm LP(2)} \rangle \approx 0.13$.

\begin{figure}
\epsfxsize=16cm
\epsfysize=16cm
\centerline{\epsffile{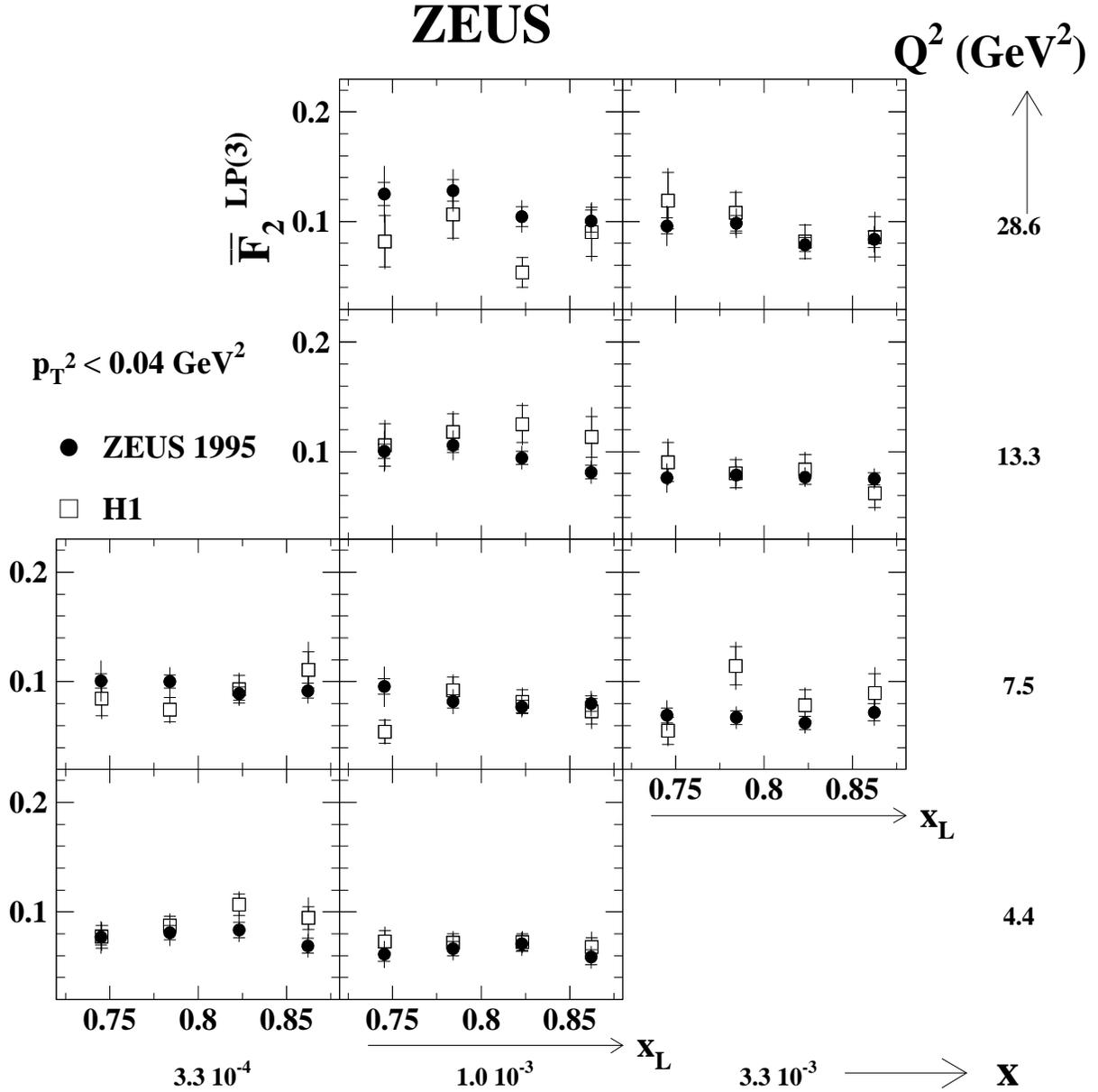}}
\vspace{-0.1cm}
\caption{The structure function $F_2^{\rm LP(3)}$ as a function of
$x_L$ in bins of $x$ and $Q^2$ (DIS sample), for protons in a restricted
$p_T^2$ range, $p_T^2< 0.04~\mathrm{GeV}^2$. 
The inner
bars show the statistical uncertainties and the outer bars are
the statistical and systematic uncertainties
added in quadrature.  The H1 results~\protect\cite{H1LP} are also shown.
}
\label{zeush1}
\end{figure}

The H1 collaboration has published~\cite{H1LP} a study of
leading proton production
for $p_T^2<0.04$~GeV$^2$, $0.7<x_L<0.9$, $2<Q^2<50$~GeV$^2$ and 
$6 \times 10^{-5} <x< 6 \times 10^{-3}$. The present analysis was 
repeated in the region of overlap with the H1 data set. A comparison 
of the H1 and ZEUS results is presented in Fig.~\ref{zeush1}. The 
agreement is good.

\subsection{Leading protons with associated dijet production}

\begin{figure}[htbp!] 
\epsfxsize=16cm
\epsfysize=16cm 
\centerline{\epsffile{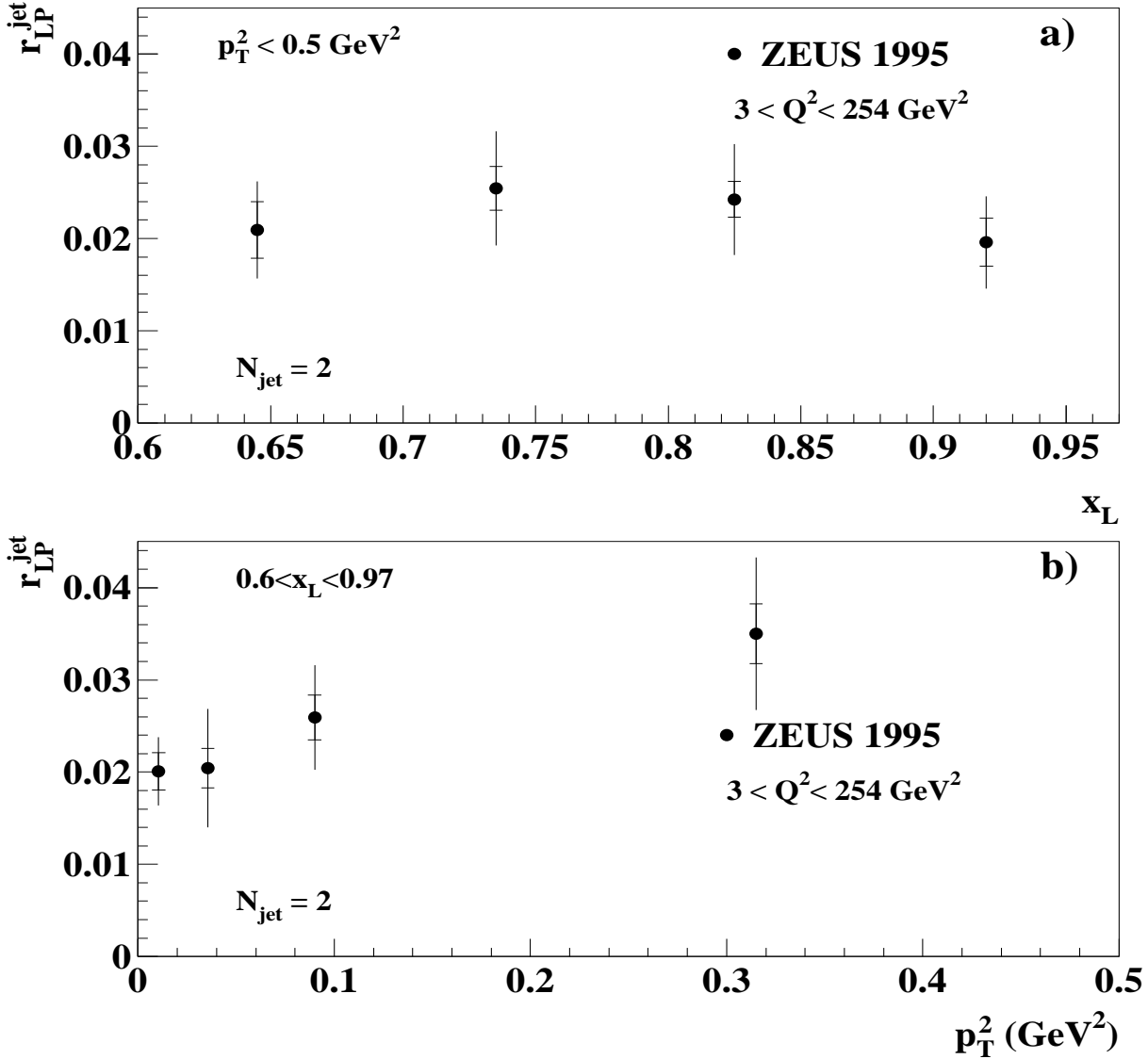}} 
\caption{Fraction of leading-proton DIS events with exactly two jets with 
$E_T> 4~\mathrm{GeV}$, $r^{\rm jet}_{\rm LP}$, as a function of a) $x_L$ 
for $p_T^2<0.5~\mathrm{GeV}^2$ and b) $p_T^2$ for $0.6<x_L<0.97$.
The inner  bars show the statistical uncertainties and the outer bars 
show the statistical and systematic uncertainties added in quadrature.
The systematic uncertainties are highly correlated. 
}
\label{jet_xl}
\end{figure}

The ratio $r^{\rm jet}_{\rm LP}$
of the yield of leading proton DIS 
events with associated dijet production to the inclusive yield of
leading proton events is presented as a function of $x_L$ and $p_T^2$ in 
Fig.~\ref{jet_xl} and Tables~\ref{tab-fig18a}-\ref{tab-fig18b}. 
The LPS 
acceptance, as well as the scattered positron 
acceptance, cancels in this ratio. The data are shown for the range 
$0.6 < x_L < 0.97$ and $p_T^2 < 0.5~\mathrm{GeV}^2$. No significant deviation 
from a flat behaviour is seen as a function of $x_L$, although there is 
some $p_T^2$ dependence. The results of this exploratory study thus 
suggest 
that the 
longitudinal- and transverse-momentum distributions of the leading proton 
are largely insensitive to 
the presence of a second hard scale, given by the transverse 
energy of the jets. 

\begin{figure}[htbp!]
\epsfxsize=17cm
\epsfysize=16cm
\centerline{\epsffile{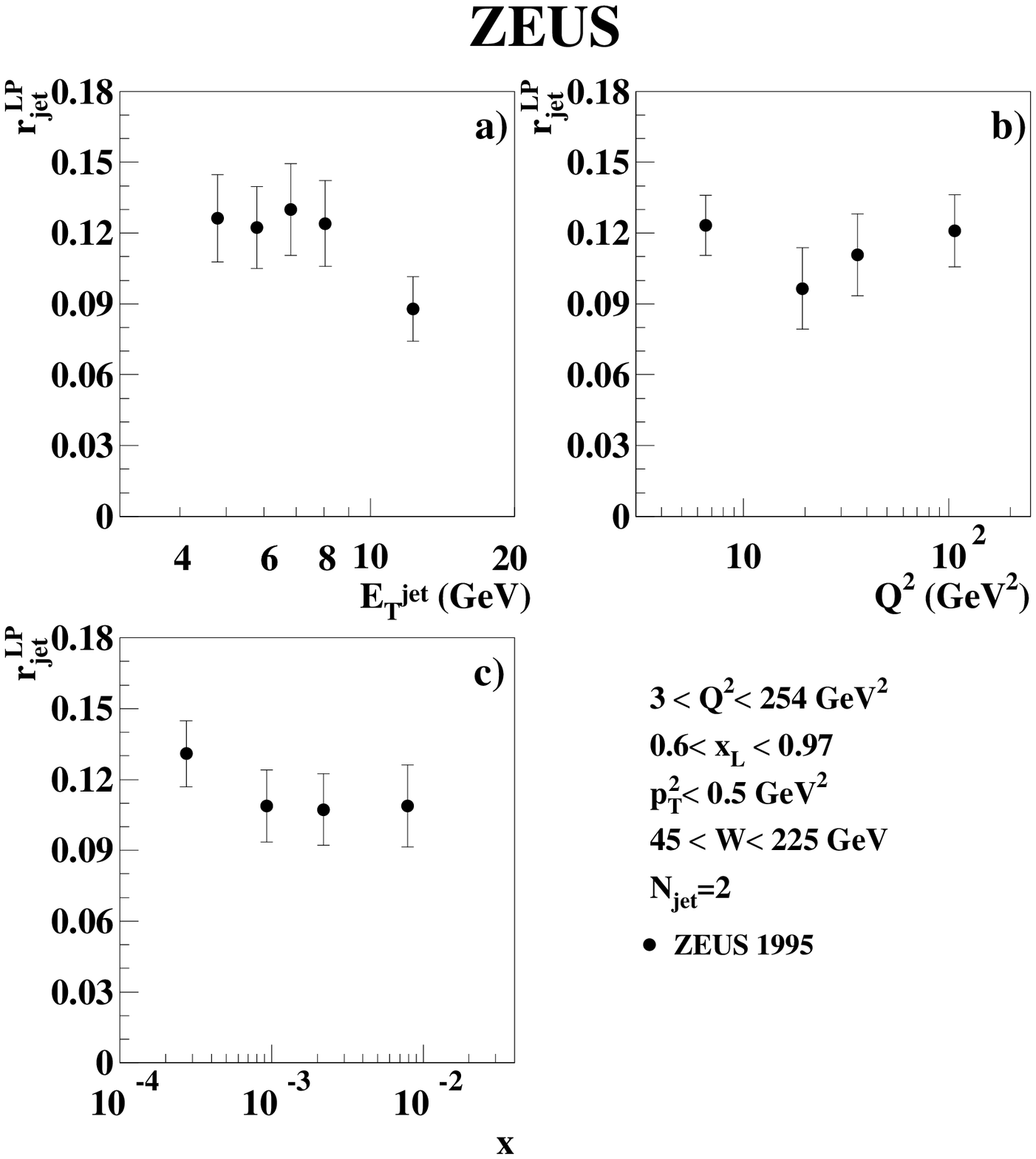}}
\caption{
  Ratio of the yield of DIS events with exactly two jets with
 $E_T> 4~\mathrm{GeV}$ and an LPS proton to the yield of DIS events with 
exactly two 
jets, also with  $E_T> 4~\mathrm{GeV}$, $r^{\rm LP}_{\rm jet}$, 
  as a function of a) $E_T$ of the higher-energy jet, b) $Q^2$ and c) $x$.  
  The error bars show the statistical
  uncertainties. A fully correlated systematic uncertainty of $\pm 13\%$
is not shown.
}
\label{morejets}
\end{figure}

Figure~\ref{morejets} and Tables~\ref{tab-fig19a}-\ref{tab-fig19c} 
present the fraction of the dijet events with a leading proton, 
$r^{\rm LP}_{\rm jet}$, plotted as a function of $E_T$, $x$ and $Q^2$.
In this case, all corrections cancel with the exception of that due to 
the LPS acceptance, which is, however, independent of $E_T$, $Q^2$ and $x$. 
The ratio is approximately independent of these variables and its value 
is consistent with that of $r^{\rm LP(2)}$. This suggests that the $E_T$,
$Q^2$ and $x$ dependences of the dijet cross section are unaffected by
the requirement of a leading proton, and that the fraction of dijet events
with a leading proton is the same as the fraction of inclusive events with
a leading proton.

\subsection{Summary}

Events of the type $e^+p \rightarrow e^+Xp$ with a final-state proton 
with $x_L>0.6$ have been studied in $e^+p$ collisions 
at HERA. The analyses used a photoproduction sample 
($Q^2<0.02$~GeV$^2$), a low-$Q^2$ sample ($0.1 <Q^2<0.74$~GeV$^2$) and
a DIS sample ($3<Q^2<254$~GeV$^2$). 

For events with a leading proton in the range $0.6<x_L<0.97$ and
$p_T^2<0.5$~GeV$^2$, the main features of the data can be summarised as
follows:

\begin{itemize}

\item  less than 10\% of the 
leading proton events in any given $x_L$ bin exhibit a large rapidity gap 
($\eta_{\max}<2.5$), indicating that diffraction is not the 
main mechanism responsible for the production of leading protons in 
this region;

\item the proton $x_L$ spectrum is only a weak function of $x_L$ for $x_L 
\ltap 0.97$;

\item the $p_T^2$ dependence of the cross section is 
well described by an exponential function, with a slope approximately 
independent of $x_L$ and $Q^2$, $b \approx 7$~GeV$^{-2}$.
The slope is also consistent with the value measured for $pp$ collisions;

\item the $x$ and $Q^2$ dependence of the semi-inclusive structure function, 
$F_2^{\rm LP}$, is similar to that of $F_2$, independently of $x_L$. 
However, $F_2^{\rm LP}$ grows with $Q^2$ slightly faster than 
$F_2$, resulting in a yield of leading protons about 20\% larger at
$Q^2=100$~GeV$^2$ than at $Q^2 \approx 0$. A similar effect was observed
for leading neutron production;

\item the shapes of the $x_L$ and $p_T^2$ spectra are largely 
unaffected by requiring two jets within the
hadronic final state $X$.  
The $Q^2$, $x$ and $E_T$ dependences 
of the dijet cross section are also broadly consistent
for leading proton events and inclusive events.

\end{itemize}

The main features of the data are reproduced by a Regge model 
assuming a superposition of Pomeron, Reggeon and pion trajectories.
The DJANGO and SCI models are ruled out by the data.

For $0.6<x_L<0.9$, the proton spectrum for (virtual-) photon-proton 
collisions is consistent with the results found in proton-proton 
reactions at lower centre-of-mass energy.
The fraction of the events with a leading proton is approximately the same 
for the $\gamma^{*}p$ and $pp$ data, in agreement with vertex 
factorisation.

In the $x_L$ region explored, a modest violation of vertex factorisation  
is observed. Nevertheless, the results of this paper indicate that the
properties of the final-state proton are largely independent of those of
the virtual photon.

\section*{Acknowledgements}

We thank the DESY Directorate for their encouragement, and
gratefully acknowledge
the support of the DESY computing and network services. We are specially
grateful to the HERA machine group: collaboration with them was crucial to
the successful installation and operation of the leading proton
spectrometer.
The design, construction  and installation of the ZEUS
detector have been made possible by the ingenuity and effort of many
people
from DESY and home institutes who are not listed as authors. Finally,
it is a pleasure to thank F.S. Navarra, N.N. Nikolaev and A. Szczurek
for many useful discussions.

\vfill\eject

\include{lp-ref}
%

\small
\begin{center}
\tablecaption{
Fraction of events with $\eta_{\max}<2.5$. 
Statistical 
uncertainties are given; 
systematic uncertainties mostly cancel in the ratio.}
\label{tab-fig5}
\begin{supertabular}{|c|c|}
\hline
$x_L$& $N_{\rm LP}(\eta_{\max}<2.5)/N_{\rm LP}$ \\
\hline 
\multicolumn{2}{|c|}{BPC}\\ 
\hline
0.63  &    0.054 $\pm$  0.030 \\
0.66  &    0.116 $\pm$  0.033 \\
0.69  &    0.075 $\pm$  0.018 \\
0.72  &    0.070 $\pm$ 0.020 \\
0.75  &    0.083 $\pm$  0.021 \\
0.78  &    0.099 $\pm$  0.020 \\
0.81  &    0.077 $\pm$ 0.018 \\
0.84  &    0.044 $\pm$  0.014 \\
0.87  &    0.066 $\pm$ 0.017 \\
0.90  &    0.042 $\pm$  0.014 \\
0.93  &    0.011 $\pm$  0.008 \\
0.96  &    0.040 $\pm$  0.016 \\
0.99  &    0.355 $\pm$  0.064 \\
1.00  &    0.812 $\pm$  0.036 \\
\hline 
\multicolumn{2}{|c|}{DIS}\\ 
\hline
0.63 & 0.0674 $\pm$ 0.0093 \\
0.66 & 0.0567 $\pm$ 0.0081 \\
0.69 & 0.0728 $\pm$ 0.0072 \\
0.72 & 0.0603 $\pm$ 0.0063 \\
0.75 & 0.0543 $\pm$ 0.0057 \\
0.78 & 0.0507 $\pm$ 0.0050 \\
0.81 & 0.0597 $\pm$ 0.0047 \\
0.84 & 0.0495 $\pm$ 0.0049 \\
0.87 & 0.0364 $\pm$ 0.0040 \\
0.90 & 0.0421 $\pm$ 0.0047 \\
0.93 & 0.0304 $\pm$ 0.0053 \\
0.96 & 0.0805 $\pm$ 0.0132 \\
0.99 & 0.3445 $\pm$ 0.0231 \\
1.00 & 0.7683 $\pm$ 0.0164 \\
\hline
\end{supertabular}
\end{center}

%
%

\newpage
\begin{center}
\tablecaption{
The normalised cross-section 
$(1/\sigma_{\rm tot}) \cdot d\sigma_{\gamma^*p \rightarrow Xp}/dx_L$ for 
the BPC and DIS data in the region $p_T^2<0.5~\mathrm{GeV}^2$. 
The two rightmost values 
indicate the statistical and systematic uncertainties, respectively.}
\label{tab-fig6}
\begin{supertabular}{|c|c|}
\hline
$x_L$ & $(1/\sigma_{\rm tot})\cdot d\sigma_{\gamma^*p \rightarrow Xp}/dx_L$ \\
\hline
\multicolumn{2}{|c|}{BPC}\\ 
\hline
0.65 &  0.339 $\pm$   0.022 $\pm$   0.050 \\
0.75 &  0.399 $\pm$   0.019 $\pm$   0.048 \\
0.83 &  0.330 $\pm$   0.014 $\pm$   0.049 \\
0.93 &  0.331 $\pm$   0.021 $\pm$   0.040 \\
0.99 &  3.60  $\pm$   0.37  $\pm$   0.54  \\
\hline
\multicolumn{2}{|c|}{DIS}\\ 
\hline
0.62 &  0.413 $\pm$   0.019 $\pm$   0.091 \\
0.65 &  0.431 $\pm$   0.017 $\pm$   0.065 \\
0.69 &  0.462 $\pm$   0.016 $\pm$   0.056 \\
0.71 &  0.445 $\pm$   0.015 $\pm$   0.066 \\
0.75 &  0.421 $\pm$   0.012 $\pm$   0.088 \\
0.77 &  0.433 $\pm$   0.012 $\pm$   0.048 \\
0.81 &  0.379 $\pm$   0.010 $\pm$   0.046 \\
0.83 &  0.359 $\pm$   0.009 $\pm$   0.029 \\
0.87 &  0.368 $\pm$   0.010 $\pm$   0.037 \\
0.89 &  0.333 $\pm$   0.010 $\pm$   0.030 \\
0.93 &  0.289 $\pm$   0.012 $\pm$   0.026 \\
0.95 &  0.46  $\pm$   0.03  $\pm$   0.11  \\
0.99 &  2.48  $\pm$   0.12  $\pm$   0.37  \\
\hline
\end{supertabular}
\end{center}

%
%

\newpage
\begin{center}
\tablecaption{
The normalised cross-section 
$(1/\sigma_{\rm tot}) \cdot d\sigma_{\gamma^*p \rightarrow Xp}/dx_L$ for 
the photoproduction, BPC and DIS data for $p_T^2 <0.04~\mathrm{GeV}^2$. 
The two rightmost values 
indicate the statistical and systematic uncertainties, respectively.}
\label{tab-fig7}
\begin{supertabular}{|c|c|}
\hline
$x_L$ & $(1/\sigma_{\rm tot})\cdot d\sigma_{\gamma^*p \rightarrow Xp}/dx_L$ \\
\hline
\multicolumn{2}{|c|}{Photoproduction}\\ 
\hline
0.64 &  0.110 $\pm$  0.017  $\pm$ 0.022 \\
0.70 &  0.081 $\pm$  0.008  $\pm$ 0.016 \\
0.76 &  0.079 $\pm$  0.006  $\pm$ 0.012 \\
0.82 &  0.090 $\pm$  0.006  $\pm$ 0.012 \\
0.88 &  0.080 $\pm$  0.006  $\pm$ 0.012 \\
\hline 
\multicolumn{2}{|c|}{BPC}\\ 
\hline
0.65 &  0.084 $\pm$  0.007 $\pm$  0.013 \\
0.75 &  0.099 $\pm$  0.005 $\pm$  0.012 \\
0.83 &  0.081 $\pm$  0.005 $\pm$  0.012 \\
0.93 & 0.0697 $\pm$ 0.0100 $\pm$ 0.0084 \\
\hline 
\multicolumn{2}{|c|}{DIS}\\ 
\hline
0.62 &  0.111 $\pm$  0.008 $\pm$  0.024 \\
0.65 &  0.105 $\pm$  0.006 $\pm$  0.016 \\
0.69 &  0.110 $\pm$  0.005 $\pm$  0.013 \\
0.71 &  0.103 $\pm$  0.004 $\pm$  0.015 \\
0.75 &  0.095 $\pm$  0.003 $\pm$  0.020 \\
0.77 &  0.102 $\pm$  0.003 $\pm$  0.011 \\
0.81 &  0.092 $\pm$  0.003 $\pm$  0.011 \\
0.83 & 0.0910 $\pm$ 0.0030 $\pm$ 0.0073 \\
0.87 & 0.0910 $\pm$ 0.0040 $\pm$ 0.0091 \\
0.89 & 0.0820 $\pm$ 0.0050 $\pm$ 0.0075 \\
\hline
\end{supertabular}
\end{center}

%
%
\newpage
\begin{center}
\tablecaption{The slopes, $b$, from fits of the functional form 
$ e^{-b p_T^2}$ to $d\sigma_{\gamma^*p \rightarrow Xp}/dp_T^2$ for leading 
protons as a function of $x_L$ for the BPC and DIS data 
samples. 
The two rightmost values 
indicate the statistical and systematic uncertainties, respectively.}
\label{tab-fig11}
\begin{supertabular}{|c|c|}
\hline
$x_L$ & $b$ (GeV$^{-2}$)\\
\hline
\multicolumn{2}{|c|}{BPC}\\ 
\hline
0.65 & 7.5 $\pm$ 1.2 $^{+0.4}_{-1.9}$ \\
0.75 & 7.6 $\pm$ 1.0 $^{+1.2}_{-1.1}$ \\
0.83 & 6.2 $\pm$ 1.2 $^{+1.2}_{-0.7}$ \\
0.93 & 4.5 $\pm$ 1.2 $^{+1.5}_{-0.8}$ \\
\hline
\multicolumn{2}{|c|}{DIS}\\ 
\hline
%
0.62 & 7.4 $\pm$ 1.2  $^{+1.0}_{-1.3}$  \\
0.65 & 7.7 $\pm$ 0.8  $^{+2.0}_{-1.2}$  \\
0.69 & 7.3 $\pm$ 0.6  $^{+0.8}_{-0.8}$  \\
0.71 & 6.8 $\pm$ 0.6  $^{+0.3}_{-0.6}$  \\
0.75 & 6.1 $\pm$ 0.6  $^{+1.2}_{-1.2}$  \\
0.77 & 6.8 $\pm$ 0.7  $^{+0.4}_{-0.4}$  \\
0.81 & 7.9 $\pm$ 0.9  $^{+0.8}_{-0.9}$  \\
0.83 & 7.3 $\pm$ 0.8  $^{+0.6}_{-0.9}$  \\
0.87 & 7.2 $\pm$ 0.7  $^{+1.1}_{-1.8}$  \\
0.89 & 6.6 $\pm$ 0.7  $^{+0.6}_{-0.5}$  \\
0.93 & 5.9 $\pm$ 0.9  $^{+1.2}_{-1.0}$  \\
0.95 & 4.4 $\pm$ 1.7  $^{+1.5}_{-0.4}$  \\
0.99 & 7.0 $\pm$ 0.9  $^{+1.9}_{-1.1}$  \\
\hline
\end{supertabular}
\end{center}

%
%
\newpage
\begin{center}
\tablecaption{
 The ratio $r^{\rm LP(3)}=\bar{F}_2^{\rm LP(3)}/F_2$ as a
function of
  $x_L$, $x$ and $Q^2$ (BPC sample), for protons with
  $p_T^2 < 0.5~\mathrm{GeV}^2$. The two rightmost values 
indicate the statistical and systematic uncertainties, respectively.}
\label{tab-fig12}
\begin{supertabular}{|c|c|c|c|}
\hline
$Q^2$ (GeV$^2)$ & $x$  & $x_L$ & $ r^{\rm LP(3)}$ \\
\hline
 0.2 & 7.4E-06 & 0.67 &  0.310 $\pm$  0.050 $\pm$  0.056 \\
 0.2 & 7.4E-06 & 0.79 &  0.408 $\pm$  0.042 $\pm$  0.049 \\
 0.2 & 7.4E-06 & 0.91 &  0.367 $\pm$  0.049 $\pm$  0.037 \\
 0.2 & 4.9E-06 & 0.67 &  0.308 $\pm$  0.052 $\pm$  0.055 \\
 0.2 & 4.9E-06 & 0.79 &  0.331 $\pm$  0.039 $\pm$  0.040 \\
 0.2 & 4.9E-06 & 0.91 &  0.361 $\pm$  0.050 $\pm$  0.036 \\
 0.2 & 3.5E-06 & 0.67 &  0.281 $\pm$  0.063 $\pm$  0.051 \\
 0.2 & 3.5E-06 & 0.79 &  0.333 $\pm$  0.049 $\pm$  0.040 \\
 0.2 & 3.5E-06 & 0.91 &  0.377 $\pm$  0.065 $\pm$  0.038 \\
 0.4 & 2.6E-05 & 0.67 &  0.357 $\pm$  0.031 $\pm$  0.064 \\
 0.4 & 2.6E-05 & 0.79 &  0.356 $\pm$  0.022 $\pm$  0.043 \\
 0.4 & 2.6E-05 & 0.91 &  0.317 $\pm$  0.026 $\pm$  0.032 \\
 0.4 & 1.3E-05 & 0.67 &  0.356 $\pm$  0.061 $\pm$  0.064 \\
 0.4 & 1.3E-05 & 0.79 &  0.405 $\pm$  0.047 $\pm$  0.049 \\
 0.4 & 1.3E-05 & 0.91 &  0.265 $\pm$  0.047 $\pm$  0.027 \\
 0.4 & 8.8E-06 & 0.67 &  0.357 $\pm$  0.084 $\pm$  0.064 \\
 0.4 & 8.8E-06 & 0.79 &  0.317 $\pm$  0.057 $\pm$  0.038 \\
 0.4 & 8.8E-06 & 0.91 &  0.298 $\pm$  0.068 $\pm$  0.030 \\
 0.6 & 4.3E-05 & 0.67 &  0.438 $\pm$  0.063 $\pm$  0.079 \\
 0.6 & 4.3E-05 & 0.79 &  0.386 $\pm$  0.043 $\pm$  0.046 \\
 0.6 & 4.3E-05 & 0.91 &  0.354 $\pm$  0.051 $\pm$  0.035 \\
\hline
\end{supertabular}
\end{center}

\newpage
\begin{center}
\tablecaption{
 The ratio $r^{\rm LP(3)}=\bar{F}_2^{\rm LP(3)}/F_2$ as a
function of
  $x_L$, $x$ and $Q^2$ (DIS sample, up to $Q^2=12~\mathrm{GeV}^2$), for 
protons 
with
  $p_T^2 < 0.5~\mathrm{GeV}^2$. The two rightmost values 
indicate the statistical and systematic uncertainties, respectively.}
\label{tab-fig13a}
\scriptsize
\begin{supertabular}{|c|c|c|c|}
\hline
$Q^2$ (GeV$^2)$ & $x$  & $x_L$ & $ r^{\rm LP(3)}$ \\
\hline
 4.0 & 1.5E-04 & 0.67 &  0.344 $\pm$  0.033  $\pm$ 0.062 \\
 4.0 & 1.5E-04 & 0.79 &  0.345 $\pm$  0.023  $\pm$ 0.041 \\
 4.0 & 1.5E-04 & 0.91 &  0.326 $\pm$  0.029  $\pm$ 0.033 \\
 4.0 & 2.5E-04 & 0.67 &  0.327 $\pm$  0.035  $\pm$ 0.059 \\
 4.0 & 2.5E-04 & 0.79 &  0.353 $\pm$   0.026 $\pm$  0.042 \\
 4.0 & 2.5E-04 & 0.91 &  0.313 $\pm$  0.031  $\pm$0.031 \\
 4.0 & 4.4E-04 & 0.67 &  0.367 $\pm$  0.035  $\pm$ 0.066 \\
 4.0 & 4.4E-04 & 0.79 &  0.335 $\pm$  0.024  $\pm$ 0.040 \\
 4.0 & 4.4E-04 & 0.91 &  0.305 $\pm$ 0.029   $\pm$0.030 \\
 4.0 & 9.8E-04 & 0.67 &  0.443 $\pm$  0.040  $\pm$ 0.080 \\
 4.0 & 9.8E-04 & 0.79 &  0.354 $\pm$  0.026  $\pm$ 0.042 \\
 4.0 & 9.8E-04 & 0.91 &  0.303 $\pm$  0.030  $\pm$ 0.030 \\
 8.0 & 2.5E-04 & 0.67 &  0.398 $\pm$  0.031  $\pm$ 0.072 \\
 8.0 & 2.5E-04 & 0.79 &  0.359 $\pm$  0.021  $\pm$ 0.043 \\
 8.0 & 2.5E-04 & 0.91 &  0.382 $\pm$  0.028  $\pm$ 0.038 \\
 8.0 & 4.4E-04 & 0.67 &  0.387 $\pm$  0.028  $\pm$ 0.070 \\
 8.0 & 4.4E-04 & 0.79 &  0.409 $\pm$  0.020  $\pm$ 0.049 \\
 8.0 & 4.4E-04 & 0.91 &  0.335 $\pm$  0.023  $\pm$ 0.033 \\
 8.0 & 9.8E-04 & 0.67 &  0.451 $\pm$  0.031  $\pm$ 0.081 \\
 8.0 & 9.8E-04 & 0.79 &  0.420 $\pm$  0.021  $\pm$ 0.050 \\
 8.0 & 9.8E-04 & 0.91 &  0.375 $\pm$  0.026  $\pm$ 0.038 \\
 8.0 & 2.2E-03 & 0.67 &  0.515 $\pm$  0.050  $\pm$ 0.093 \\
 8.0 & 2.2E-03 & 0.79 &  0.380 $\pm$  0.031  $\pm$ 0.046 \\
 8.0 & 2.2E-03 & 0.91 &  0.421 $\pm$  0.041  $\pm$ 0.042 \\
 12.0 & 4.4E-04 & 0.67 &  0.355$\pm$  0.029  $\pm$ 0.064 \\
 12.0 & 4.4E-04 & 0.79 &  0.379 $\pm$  0.022 $\pm$ 0.046 \\
 12.0 & 4.4E-04 & 0.91 &  0.396 $\pm$  0.028 $\pm$  0.040 \\
 12.0 & 9.8E-04 & 0.67 &  0.416 $\pm$  0.029 $\pm$  0.075 \\
 12.0 & 9.8E-04 & 0.79 &  0.387 $\pm$  0.020 $\pm$  0.046 \\
 12.0 & 9.8E-04 & 0.91 &  0.369 $\pm$  0.025 $\pm$  0.037 \\
 12.0 & 2.2E-03 & 0.67 &  0.437 $\pm$  0.039 $\pm$  0.079 \\
 12.0 & 2.2E-03 & 0.79 &  0.407 $\pm$  0.027 $\pm$  0.049 \\
 12.0 & 2.2E-03 & 0.91 &  0.445 $\pm$  0.036 $\pm$  0.044 \\
 12.0 & 4.0E-03 & 0.67 &  0.462 $\pm$  0.067 $\pm$  0.083 \\
 12.0 & 4.0E-03 & 0.79 &  0.427 $\pm$  0.046 $\pm$  0.051 \\
 12.0 & 4.0E-03 & 0.91 &  0.349 $\pm$  0.053 $\pm$  0.035 \\
\hline
\end{supertabular}
\end{center}

\normalsize

\newpage
\begin{center}
\tablecaption{
 The ratio $r^{\rm LP(3)}=\bar{F}_2^{\rm LP(3)}/F_2$ as a
function of
  $x_L$, $x$ and $Q^2$ (DIS sample, for $Q^2> 12~\mathrm{GeV}^2$), for 
protons 
with
  $p_T^2 < 0.5~\mathrm{GeV}^2$. The two rightmost values 
indicate the statistical and systematic uncertainties, respectively.}
\label{tab-fig13b}
\footnotesize
\begin{supertabular}{|c|c|c|c|}
\hline
$Q^2$ (GeV$^2)$ & $x$  & $x_L$ & $ r^{\rm LP(3)}$ \\
\hline
 21.0 & 9.8E-04 & 0.67 &  0.423 $\pm$  0.027 $\pm$  0.076 \\
 21.0 & 9.8E-04 & 0.79 &  0.387 $\pm$  0.019 $\pm$  0.046 \\
 21.0 & 9.8E-04 & 0.91 &  0.360 $\pm$  0.023 $\pm$  0.036 \\
 21.0 & 2.2E-03 & 0.67 &  0.437 $\pm$  0.034 $\pm$  0.079 \\
 21.0 & 2.2E-03 & 0.79 &  0.427 $\pm$  0.024 $\pm$  0.051 \\
 21.0 & 2.2E-03 & 0.91 &  0.379 $\pm$  0.029 $\pm$  0.038 \\
 21.0 & 4.0E-03 & 0.67 &  0.486 $\pm$  0.049 $\pm$  0.087 \\
 21.0 & 4.0E-03 & 0.79 &  0.390 $\pm$  0.031 $\pm$  0.047 \\
 21.0 & 4.0E-03 & 0.91 &  0.361 $\pm$ 0.038  $\pm$ 0.036 \\
 21.0 & 5.9E-03 & 0.67 &  0.474 $\pm$  0.066 $\pm$  0.085 \\
 21.0 & 5.9E-03 & 0.79 &  0.372 $\pm$  0.042  $\pm$ 0.045 \\
 21.0 & 5.9E-03 & 0.91 &  0.466 $\pm$  0.059  $\pm$ 0.047 \\
 46.0 & 2.2E-03 & 0.67 &  0.452 $\pm$  0.043  $\pm$ 0.081 \\
 46.0 & 2.2E-03 & 0.79 &  0.372 $\pm$  0.028  $\pm$ 0.045 \\
 46.0 & 2.2E-03 & 0.91 &  0.490 $\pm$  0.040  $\pm$ 0.049 \\
 46.0 & 4.0E-03 & 0.67 &  0.444 $\pm$  0.055  $\pm$ 0.080 \\
 46.0 & 4.0E-03 & 0.79 &  0.433 $\pm$  0.039  $\pm$ 0.052 \\
 46.0 & 4.0E-03 & 0.91 &  0.360 $\pm$  0.045  $\pm$ 0.036 \\
 46.0 & 5.9E-03 & 0.67 &  0.435 $\pm$  0.064  $\pm$ 0.078 \\
 46.0 & 5.9E-03 & 0.79 &  0.348 $\pm$  0.041  $\pm$ 0.042 \\
 46.0 & 5.9E-03 & 0.91 &  0.436 $\pm$ 0.059   $\pm$0.044 \\
 46.0 & 1.0E-02 & 0.67 &  0.453 $\pm$  0.050  $\pm$ 0.081 \\
 46.0 & 1.0E-02 & 0.79 &  0.412 $\pm$  0.034  $\pm$ 0.049 \\
 46.0 & 1.0E-02 & 0.91 &  0.409 $\pm$  0.043  $\pm$ 0.041 \\
 130.0 & 5.9E-03 & 0.67 &  0.441 $\pm$  0.094 $\pm$ 0.079 \\
 130.0 & 5.9E-03 & 0.79 &  0.327 $\pm$  0.058 $\pm$  0.039 \\
 130.0 & 5.9E-03 & 0.91 &  0.528 $\pm$  0.094 $\pm$  0.053 \\
 130.0 & 1.0E-02 & 0.67 &  0.408 $\pm$  0.060 $\pm$  0.073 \\
 130.0 & 1.0E-02 & 0.79 &  0.505 $\pm$  0.048 $\pm$  0.061 \\
 130.0 & 1.0E-02 & 0.91 &  0.358 $\pm$  0.051 $\pm$  0.036 \\
 130.0 & 2.5E-02 & 0.67 &  0.421 $\pm$  0.069 $\pm$  0.076 \\
 130.0 & 2.5E-02 & 0.79 &  0.430 $\pm$  0.050 $\pm$  0.052 \\
 130.0 & 2.5E-02 & 0.91 &  0.403 $\pm$  0.062 $\pm$  0.040 \\
\hline
\end{supertabular}
\end{center}

\newpage
%
%
\begin{center}
\tablecaption{
The ratio $r^{\rm LP(2)}=\bar{F}_2^{\rm LP(2)}/F_2$
  as a function of $x$ for fixed $Q^2$ values, for protons with 
$0.6 < x_L < 0.97$ and $p_T^2 < 0.5~\mathrm{GeV}^2$. The statistical 
uncertainty is given. 
A fully correlated systematic uncertainty of $\pm 13\%$ is not included.
}
\label{tab-fig14}
\footnotesize
\begin{supertabular}{|c|c|c|}
\hline
$Q^2$ (GeV$^2)$ & $x$  & $ r^{\rm LP(2)}$ \\
\hline
 0.20 & 3.5E-06 & 0.118   $\pm$ 0.012   \\
 0.20 & 4.9E-06 & 0.1177  $\pm$  0.0093  \\
 0.20 & 7.4E-06 & 0.1309  $\pm$ 0.0095 \\
 0.36 & 8.8E-06 & 0.113   $\pm$ 0.014   \\
 0.36 & 1.3E-05 & 0.124   $\pm$ 0.010   \\
 0.36 & 2.6E-05 & 0.1211  $\pm$ 0.0092  \\
 0.60 & 4.3E-05 & 0.137   $\pm$ 0.010   \\
  4.0 & 9.8E-04 & 0.1281 $\pm$ 0.0063 \\
  4.0 & 4.4E-04 & 0.1186 $\pm$ 0.0058 \\
  4.0 & 2.5E-04 & 0.1190 $\pm$ 0.0062 \\
  4.0 & 1.5E-04 & 0.1206 $\pm$ 0.0056 \\
  8.0 & 2.2E-03 & 0.1507 $\pm$ 0.0079 \\
  8.0 & 9.8E-04 & 0.1473 $\pm$ 0.0052 \\
  8.0 & 4.4E-04 & 0.1358 $\pm$ 0.0048 \\
  8.0 & 2.5E-04 & 0.1332 $\pm$ 0.0053 \\
 12.0 & 4.0E-03 & 0.147  $\pm$ 0.011  \\
 12.0 & 2.2E-03 & 0.1511 $\pm$ 0.0067 \\
 12.0 & 9.8E-04 & 0.1382 $\pm$ 0.0049 \\
 12.0 & 4.4E-04 & 0.1345 $\pm$ 0.0053 \\
 21.0 & 5.9E-03 & 0.151  $\pm$ 0.011  \\
 21.0 & 4.0E-03 & 0.1438 $\pm$ 0.0077 \\
 21.0 & 2.2E-03 & 0.1477 $\pm$ 0.0058 \\
 21.0 & 9.8E-04 & 0.1377 $\pm$ 0.0046 \\
 46.0 & 1.0E-02 & 0.1496 $\pm$ 0.0084 \\
 46.0 & 5.9E-03 & 0.140  $\pm$ 0.011  \\
 46.0 & 4.0E-03 & 0.1472 $\pm$ 0.0092 \\
 46.0 & 2.2E-03 & 0.1512 $\pm$ 0.0072 \\
130.0 & 2.5E-02 & 0.149  $\pm$ 0.012  \\
130.0 & 1.0E-02 & 0.156  $\pm$ 0.011  \\
130.0 & 5.9E-03 & 0.147  $\pm$ 0.016  \\
\hline
\end{supertabular}
\end{center}

\newpage
%
%
\begin{center}
\tablecaption{
The average ratio $\langle{r}^{\rm LP(2)}\rangle=\bar{F}_2^{\rm 
LP(2)}/F_2$ as a function of $Q^2$ for $0.6<x_L<0.97$ and $p_T^2 < 
0.5~\mathrm{GeV}^2$.
The statistical  uncertainty is given. A fully correlated 
systematic uncertainty of 13\% is not included. }
\label{tab-fig15a}
\begin{supertabular}{|c|c|}
\hline
$Q^2$ (GeV$^2)$ &  $ \langle r^{\rm LP(2)} \rangle$ \\
\hline
  0.29 & 0.1230 $\pm$ 0.0033 \\
  5.21 & 0.1312 $\pm$ 0.0020 \\
 16.60 & 0.1397 $\pm$ 0.0021 \\
 69.00 & 0.1471 $\pm$ 0.0036 \\
\hline
\end{supertabular}
\end{center}

\vspace{5cm}

%
\begin{center}
\tablecaption{
The average ratio $\langle{r}^{\rm LP(2)}\rangle$ as
a function of $Q^2$ for two different $p_T^2$ ranges normalised to the
value at $Q^2=0.25~\mathrm{GeV}^2$. The 
statistical
uncertainty is given; systematic errors mostly cancel in the ratio.}
\label{tab-fig15b}
\begin{supertabular}{|c|c|c|}
\hline
$Q^2$ (GeV$^2)$ & \multicolumn{2}{ c|} { $ \langle r^{\rm 
LP(2)}(Q^2)\rangle/  
\langle r^{\rm LP(2)}(Q^2=0.25~{\mbox {GeV}}^2 \rangle $)}  \\
\hline
        &        $p_T^2<0.04~\mathrm{GeV}^2$ & $p_T^2<0.5~\mathrm{GeV}^2$\\
\hline
  0.002 &  0.941 $\pm$  0.033 &                \\
  0.29  &  1.000 $\pm$  0.000 &  1.000 $\pm$  0.000 \\
  5.21  &  1.062 $\pm$  0.022 &  1.067 $\pm$  0.031 \\
 16.60  &  1.115 $\pm$  0.023 &  1.136 $\pm$  0.031 \\
 69.00  &  1.152 $\pm$  0.039 &  1.196 $\pm$  0.037 \\
\hline
\end{supertabular}
\end{center}

\newpage

%
%
\begin{center}
\tablecaption{
The structure-function $\bar{F}_2^{\rm LP(2)}$ as a function of $x$
for
$0.6<x_L<0.97$ and $p_T^2<0.5~\mathrm{GeV}^2$. The statistical
uncertainty is given. A fully 
correlated systematic uncertainty of $\pm 13\%$ is not included, nor is 
the uncertainty of the $F_2$ parametrisations used.
}
\label{tab-fig16}
\footnotesize
\begin{supertabular}{|c|c|c|}
\hline
   $x$   & $Q^2$ (GeV$^2)$ & $\bar{F}_2^{\rm LP(2)}$ \\
\hline
 3.5E-06 &   0.2 & 0.0286 $\pm$  0.0029 \\
 4.9E-06 &   0.2 & 0.0277 $\pm$ 0.0022 \\
 7.4E-06 &   0.2 & 0.0296 $\pm$ 0.0021 \\
 8.8E-06 &   0.4 & 0.0391 $\pm$ 0.0048 \\
 1.3E-05 &   0.4 & 0.0412 $\pm$ 0.0035 \\
 2.6E-05 &   0.4 & 0.0378 $\pm$ 0.0029 \\
 4.3E-05 &   0.6 & 0.0559 $\pm$ 0.0042 \\
 1.5E-04 &   4.0 & 0.1126 $\pm$ 0.0052 \\
 2.5E-04 &   4.0 & 0.0991 $\pm$ 0.0052 \\
 4.4E-04 &   4.0 & 0.0872 $\pm$ 0.0043 \\
 9.8E-04 &   4.0 & 0.0793 $\pm$ 0.0039 \\
 2.5E-04 &   8.0 & 0.1518 $\pm$ 0.0060 \\
 4.4E-04 &   8.0 & 0.1347 $\pm$ 0.0048 \\
 9.8E-04 &   8.0 & 0.1201 $\pm$ 0.0042 \\
 2.2E-03 &   8.0 & 0.1014 $\pm$ 0.0053 \\
 4.4E-04 &  12.0 & 0.1563 $\pm$ 0.0062 \\
 9.8E-04 &  12.0 & 0.1303 $\pm$ 0.0046 \\
 2.2E-03 &  12.0 & 0.1157 $\pm$ 0.0051 \\
 4.0E-03 &  12.0 & 0.0969 $\pm$ 0.0073 \\
 9.8E-04 &  21.0 & 0.1552 $\pm$ 0.0052 \\
 2.2E-03 &  21.0 & 0.1325 $\pm$ 0.0052 \\
 4.0E-03 &  21.0 & 0.1093 $\pm$ 0.0059 \\
 5.9E-03 &  21.0 & 0.1029 $\pm$ 0.0074 \\
 2.2E-03 &  46.0 & 0.1659 $\pm$ 0.0079 \\
 4.0E-03 &  46.0 & 0.1339 $\pm$ 0.0084 \\
 5.9E-03 &  46.0 & 0.1127 $\pm$ 0.0086 \\
 1.0E-02 &  46.0 & 0.1026 $\pm$ 0.0058 \\
 5.9E-03 & 130.0 & 0.140  $\pm$ 0.015  \\
 1.0E-02 & 130.0 & 0.1240 $\pm$ 0.0086 \\
 2.5E-02 & 130.0 & 0.0880 $\pm$ 0.0071 \\
\hline
\end{supertabular}
\end{center}

\newpage

%
%
%
\begin{center}
\tablecaption{
The structure function $F_2^{\rm LP(3)}$ as a function of
$x_L$ in bins of $x$ and $Q^2$ (DIS sample), for protons in a restricted
$p_T^2$ range, $p_T^2< 0.04~\mathrm{GeV}^2$.
The two rightmost values 
indicate the statistical and systematic uncertainties, respectively. }
\label{tab-fig17}
\scriptsize
\begin{supertabular}{|c|c|c|c|}
\hline
$Q^2$ (GeV$^2)$ &  $x$    &  $x_L$ &    $\bar{F}_2^{\rm LP(3)}$        \\
\hline
 4.4 & 3.3E-04 &   0.73 & 0.0675 $\pm$  0.0080 $\pm$ 0.0157 \\
 4.4 & 3.3E-04 &   0.78 & 0.0724 $\pm$ 0.0078 $\pm$ 0.0116 \\
 4.4 & 3.3E-04 &   0.83 & 0.0756 $\pm$ 0.0082 $\pm$ 0.0106 \\
 4.4 & 3.3E-04 &   0.88 & 0.0585 $\pm$ 0.0082 $\pm$ 0.0098 \\
 4.4 & 1.0E-03 &   0.73 & 0.0489 $\pm$ 0.0077 $\pm$ 0.0125 \\
 4.4 & 1.0E-03 &   0.78 & 0.0552 $\pm$ 0.0077 $\pm$ 0.0101 \\
 4.4 & 1.0E-03 &   0.83 & 0.0605 $\pm$ 0.0083 $\pm$ 0.0099 \\
 4.4 & 1.0E-03 &   0.88 & 0.0460 $\pm$ 0.0083 $\pm$ 0.0093 \\
 7.5 & 3.3E-04 &   0.73 & 0.0962 $\pm$ 0.0079 $\pm$ 0.0209 \\
 7.5 & 3.3E-04 &   0.78 & 0.0955 $\pm$ 0.0074 $\pm$ 0.0136 \\
 7.5 & 3.3E-04 &   0.83 & 0.0822 $\pm$ 0.0070 $\pm$ 0.0101 \\
 7.5 & 3.3E-04 &   0.88 & 0.0855 $\pm$ 0.0083 $\pm$ 0.0115 \\
 7.5 & 1.0E-03 &   0.73 & 0.0899 $\pm$ 0.0085 $\pm$ 0.0200 \\
 7.5 & 1.0E-03 &   0.78 & 0.0737 $\pm$ 0.0072 $\pm$ 0.0114 \\
 7.5 & 1.0E-03 &   0.83 & 0.0682 $\pm$ 0.0071 $\pm$ 0.0094 \\
 7.5 & 1.0E-03 &   0.88 & 0.0715 $\pm$ 0.0084 $\pm$ 0.0107 \\
 7.5 & 3.3E-03 &   0.73 & 0.0587 $\pm$ 0.0081 $\pm$ 0.0143 \\
 7.5 & 3.3E-03 &   0.78 & 0.0563 $\pm$ 0.0075 $\pm$ 0.0100 \\
 7.5 & 3.3E-03 &   0.83 & 0.0505 $\pm$ 0.0072 $\pm$ 0.0085 \\
 7.5 & 3.3E-03 &   0.88 & 0.0619 $\pm$ 0.0093 $\pm$ 0.0109 \\
 13.3 & 1.0E-03 &   0.73 & 0.0957 $\pm$  0.0078 $\pm$ 0.0208 \\
 13.3 & 1.0E-03 &   0.78 & 0.1023 $\pm$ 0.0076 $\pm$ 0.0144 \\
 13.3 & 1.0E-03 &   0.83 & 0.0886 $\pm$ 0.0072 $\pm$ 0.0107 \\
 13.3 & 1.0E-03 &   0.88 & 0.0729 $\pm$ 0.0074 $\pm$ 0.0101 \\
 13.3 & 3.3E-03 &   0.73 & 0.0666 $\pm$ 0.0061 $\pm$ 0.0147 \\
 13.3 & 3.3E-03 &   0.78 & 0.0698 $\pm$ 0.0059 $\pm$ 0.0102 \\
 13.3 & 3.3E-03 &   0.83 & 0.0677 $\pm$ 0.0059 $\pm$ 0.0084 \\
 13.3 & 3.3E-03 &   0.88 & 0.0658 $\pm$ 0.0067 $\pm$ 0.0091 \\
 28.6 & 1.0E-03 &   0.73 & 0.1249 $\pm$ 0.0125 $\pm$ 0.0281 \\
 28.6 & 1.0E-03 &   0.78 & 0.1290 $\pm$ 0.0119 $\pm$ 0.0195 \\
 28.6 & 1.0E-03 &   0.83 & 0.1007 $\pm$ 0.0107 $\pm$ 0.0140 \\
 28.6 & 1.0E-03 &   0.88 & 0.0956 $\pm$ 0.0121 $\pm$ 0.0150 \\
 28.6 & 3.3E-03 &   0.73 & 0.0906 $\pm$ 0.0089 $\pm$ 0.0203 \\
 28.6 & 3.3E-03 &   0.78 & 0.0934 $\pm$ 0.0085 $\pm$ 0.0140 \\
 28.6 & 3.3E-03 &   0.83 & 0.0701 $\pm$ 0.0075 $\pm$ 0.0097 \\
 28.6 & 3.3E-03 &   0.88 & 0.0758 $\pm$ 0.0089 $\pm$ 0.0114 \\
\hline
\end{supertabular}
\end{center}

\newpage

\begin{center}
\tablecaption{
Fraction of leading-proton DIS events with exactly two jets with
$E_T> 4~\mathrm{GeV}$, $r^{\rm jet}_{\rm LP}$, as a function of $x_L$ for
$p_T^2<0.5~\mathrm{GeV}^2$. The two rightmost values 
indicate the statistical and systematic uncertainties, respectively. 
The systematic uncertainties are highly correlated.
}
\label{tab-fig18a}
\begin{supertabular}{|c|c|}
\hline
$x_L$  &  $r^{\rm jet}_{\rm LP}$ \\
\hline
 0.645 &  0.0209 $\pm$ 0.0031 $\pm$ 0.0053\\
 0.735 &  0.0254 $\pm$ 0.0023 $\pm$ 0.0062\\
 0.825 &  0.0242 $\pm$ 0.0019 $\pm$ 0.0060 \\
 0.920 &  0.0196 $\pm$ 0.0026 $\pm$ 0.0050\\
\hline
\end{supertabular}
\end{center}

\vspace{5cm}

\begin{center}
\tablecaption{
Fraction of leading-proton DIS events with exactly two jets with
$E_T> 4~\mathrm{GeV}$, $r^{\rm jet}_{\rm LP}$, as a function of $p_T^2$ 
for $0.6 <x_L<0.97$.  The two rightmost values 
indicate the statistical and systematic uncertainties, respectively. 
The systematic uncertainties are highly correlated.
}
\label{tab-fig18b}
\begin{supertabular}{|c|c|}
\hline
$p_T^2$ (GeV$^{2})$ &  $r^{\rm jet}_{\rm LP}$ \\
\hline
 0.0105 & 0.0201 $\pm$ 0.0020 $\pm$ 0.0037\\
 0.0355 & 0.0204 $\pm$ 0.0022 $\pm$ 0.0064\\
 0.0900 & 0.0259 $\pm$ 0.0024 $\pm$ 0.0057\\
 0.3150 & 0.0350 $\pm$ 0.0032 $\pm$ 0.0083\\
\hline
\end{supertabular}
\end{center}

\newpage

\begin{center}
\tablecaption{
 Ratio of the yield of DIS events with exactly two jets with
 $E_T> 4~\mathrm{GeV}$ and an LPS proton to the yield of DIS events with
 exactly two jets, also with  $E_T> 4~\mathrm{GeV}$, $r^{\rm LP}_{\rm jet}$,
 as a function of $E_T$ of the higher-energy jet. The statistical 
uncertainty is given. A fully correlated systematic uncertainty of $\pm
13\%$ is 
not included.} 
\label{tab-fig19a}
\begin{supertabular}{|c|c|}
\hline
$E_T$ (GeV) &  $r^{\rm LP}_{\rm jet}$ \\
\hline
4.8  & 0.126    $\pm$  0.018 \\
5.8  & 0.122    $\pm$  0.017 \\
6.8  & 0.130    $\pm$  0.019 \\
8.0  & 0.124    $\pm$  0.018 \\
12.3 & 0.088    $\pm$  0.014 \\
\hline
\end{supertabular}
\end{center}

\vspace{1.5 cm}

\begin{center}
\tablecaption{
 Ratio of the yield of DIS events with exactly two jets with
 $E_T> 4~\mathrm{GeV}$ and an LPS proton to the yield of DIS events with
 exactly two jets, also with  $E_T> 4~\mathrm{GeV}$, $r^{\rm LP}_{\rm jet}$,
 as a function of $Q^2$. The statistical 
uncertainty is given. A fully correlated systematic uncertainty of $\pm
13\%$ is 
not included.} 
\label{tab-fig19b}
\begin{supertabular}{|c|c|}
\hline
$Q^2$ (GeV$^2$) & $r^{\rm LP}_{\rm jet}$ \\
\hline
 6.6   & 0.123  $\pm$ 0.013  \\
 19.4  & 0.097  $\pm$ 0.017 \\
 36.0  & 0.111  $\pm$ 0.017 \\
 106.8 & 0.121  $\pm$ 0.015  \\
\hline
\end{supertabular}
\end{center}

\vspace{1.5 cm}

\begin{center}
\tablecaption{
 Ratio of the yield of DIS events with exactly two jets with
 $E_T> 4~\mathrm{GeV}$ and an LPS proton to the yield of DIS events with
 exactly two jets, also with  $E_T> 4~\mathrm{GeV}$, $r^{\rm LP}_{\rm jet}$,
 as a function of $x$. The statistical 
uncertainty is given. A fully correlated systematic uncertainty of $\pm
13\%$ is 
not included.} 
\label{tab-fig19c}
\begin{supertabular}{|c|c|}
\hline
$x$       & $r^{\rm LP}_{\rm jet}$ \\
\hline
 0.00027  &  0.131  $\pm$ 0.014   \\
 0.00093  &  0.109  $\pm$ 0.015  \\
 0.0022   &  0.107  $\pm$ 0.015 \\
 0.0079   &  0.109  $\pm$ 0.017  \\
\hline
\end{supertabular}
\end{center}


\clearpage

\clearpage

\clearpage
\newpage

\clearpage
\newpage

\clearpage

\clearpage
\newpage

\clearpage

%
%

\begin{thebibliography}{99}

\bibitem{review1} See e.g.:\\
G. Alberi and G. Goggi, Phys. Rep. {\bf 74}, 1 (1981);\\
K. Goulianos, Phys. Rep. {\bf 102}, 169 (1983), and references therein.

\bibitem{review2} 
M. Basile et al., Nuovo Cimento {\bf 66A}, 129  (1981).

\bibitem{review2a} 
M. Basile et al., Lettere al Nuovo Cimento {\bf 32}, 321 (1981);\\
V.N. Gribov et al., {\it The creation of QCD and the effective energy},
 L.N. Lipatov (ed.), World Scientific Series in 20th Century Physics, Vol. 
25, World Scientific, Singapore (2001).

\bibitem{FNC}
ZEUS Collab., M.\ Derrick et al., Phys. Lett. {\bf B 384}, 388 (1996).

\bibitem{H1LP}
H1 Collab., C. Adloff et al., \EP{C 6}{1999}{587}.

\bibitem{zeuslndijet}
ZEUS Collab., J.\ Breitweg et al.,
\NP{B 593}{2000}{1}.

\bibitem{h1php}H1 Collab., C. Adloff et al., Nucl. Phys. {\bf B 619}, 3 
(2001).

\bibitem{neutrons}  ZEUS Collab., S. Chekanov et al., 
Nucl. Phys. {\bf B 637}, 3 (2002).

\bibitem{ISR} G. Bellettini et al., \PL{B 45}{1973}{69};\\
G.J. Alner et al., Z. Phys. {\bf C 33}, 1 (1986).

\bibitem{vdm} J.J. Sakurai, Ann. Phys. (NY) {\bf 11}, 1 (1960);\\
J.J. Sakurai, \PRL{22}{1969}{981}.

\bibitem{dalesio} N.N. Nikolaev, J. Speth and B.G. Zakharov,
hep-ph/9708290 (1997);\\
U. D'Alesio and H.J. Pirner, \EP{A 7}{2000}{109}.

\bibitem{SULLIVAN}
J.\ D.\ Sullivan, \PR{D 5}{1972}{1732}.

\bibitem{ZOLLER}
R.\ Zoller, \ZP{C 53}{1992}{443}.

\bibitem{HOLTMANN}
H.\ Holtmann et al., \PL{B 338}{1994}{363}.

\bibitem{KOPE}
B.\ Kopeliovich, B.\ Povh and I.\ Potashnikova, \ZP{C 73}{1996}{125}.

\bibitem{SNS}
A.\ Szczurek, N.N.\ Nikolaev and J.\ Speth, \PL{B 428}{1998}{383}.


\bibitem{sci}
A.\ Edin, G.\ Ingelman and J.\ Rathsman, \PL{B 366}{1996}{371}.

\bibitem{FF}
L. Trentadue and G. Veneziano, \PL{B 323}{1994}{210};\\
D. Graudenz, \NP{B 432}{1994}{351};\\
L.\ Trentadue, \NPPS{39 BC}{1995}{50};\\
L.\ Trentadue, \NPPS{64}{1998}{152}.\\
For applications to HERA see also: \\
D.\ de Florian and R.\ Sassot, \PR {D 56}{1997}{426} and 
\PR{D 58}{1998}{054003}.

\bibitem{lpsrho} ZEUS Collab., M.\ Derrick et al., \ZP{C 73}{1997}{253}.

\bibitem{bluebook} ZEUS Collab., U. Holm (ed.),
{\it The ZEUS Detector}, Status Report (unpublished)
DESY, 1993, available on\\ 
\verb+http://www-zeus.desy.de/bluebook/bluebook.html+.

\bibitem{ctd} N. Harnew et al., \NIM{A 279}{1989}{290};\\
B. Foster et al., Nucl. Phys. Proc. Suppl. {\bf B 32}, 181 (1993);\\
B. Foster et al., \NIM{A 338}{1994}{254}.

\bibitem{cal} M. Derrick et al., \NIM{A 309}{1991}{77};\\
A. Andresen et al., \NIM{A 309}{1991}{101};\\
A. Caldwell et al., \NIM{A 321}{1992}{356};\\
A. Bernstein et al., \NIM{A 336}{1993}{23}.

\bibitem{lumi} J. Andruszk\'ow et al., Report DESY-92-066, DESY, 1992;\\
ZEUS Collab., M. Derrick et al., \ZP{C 63}{1994}{391};\\
 J. Andruszk\'ow et al., Acta Phys. Pol. {\bf B 32}, 2025 (2001).

\bibitem{zeusbpc}
ZEUS Collab., J. Breitweg et al., \PL{B 407}{1997}{432}.

\bibitem{zeusbpt}
ZEUS Collab., J. Breitweg et al., \PL{B 487}{2000}{53}.

\bibitem{surrow} B. Surrow, 
Ph.D. thesis, University of Hamburg (1998), 
Report DESY-THESIS-1998-004.

\bibitem{monteiro} T. Monteiro, Ph.D. thesis, University of Hamburg (1998), 
Report DESY-THESIS-1998-027.


\bibitem{srtd} A. Bamberger et al., \NIM{A 401}{1997}{63}.

\bibitem{DA}
S. Bentvelsen, J. Engelen and P. Kooijman, 
{\it Proceedings of the
Workshop on Physics at HERA}, Volume~1, W. Buchm\"uller and G. Ingelman
(eds.), DESY (1991) p.~23;\\
K. C. Hoeger, ibid., p.~43.

\bibitem{zeusdiff94} ZEUS Collab., J. Breitweg et al., \EP{C 6}
{1999}{43}.


\bibitem{gb} G.M. Briskin, Ph.D. Thesis, University of Tel Aviv
(1998), Report DESY-THESIS-1998-036.

\bibitem{jacblo}
F. Jacquet and A. Blondel, {\it Proceedings of the Study for an $ep$
Facility for Europe}, U. Amaldi (Ed.), p.~391, Hamburg, Germany (1979).
Also Preprint DESY 79/48. 

\bibitem{piontoproton} 
P. Capiluppi et al., \NP{B 70}{1974}{1};\\
P. Capiluppi et al., \NP{B 79}{1974}{189};\\
M.G. Albrow et al., \NP{B 73}{1974}{40}.


\bibitem{tesi_yuri} Y. Garcia-Zamora, Ph.D. thesis, University of Geneva
(1998) (unpublished).

\bibitem{t_lps} ZEUS Collab., J. Breitweg et al., \EP{C 2}
{1998}{247}.

\bibitem{tesi_alberto} A. Garfagnini, Ph.D. thesis, University of Calabria
(1998) (unpublished).

\bibitem{review_vm} See e.g.:\\
J.A. Crittenden, {\it Exclusive Production of Neutral Vector Mesons at the
Electron-Proton Collider HERA}, Springer Tracts in Modern Physics, Vol.
140, Springer, Berlin, Germany (1997), and references therein;\\
H. Abramowicz and A. Caldwell, Rev. Mod. Phys. {\bf 71}, 1275 (1999), and
references therein.

\bibitem{tesi_mariacarmela} M.C. Petrucci, Ph.D. thesis, University of
Torino (1999), unpublished.


\bibitem{JETKT} S. Catani et al., \PL{B 269}{1991}{432}.

\bibitem{mx} ZEUS Collab., J. Breitweg et al., \ZP{C 75}{1997}{421}.

\bibitem{michal}M. Kasprzak, Ph.D. thesis, University of Warsaw (1996),
Internal Report DESY-F35D-96-16.

\bibitem{thesis_masahide}M. Inuzuka, Ph.D. thesis, University of Tokyo
(1999), KEK Report 99-9.

\bibitem{heracles} K. Kwiatkowski, H. Spiesberger and H.-J. M\"ohring, \CPC{69}{1992}{155}. 

\bibitem{rg}H.\ Jung, \CPC{86}{1995}{147}.

\bibitem{thesis_peter}
C.-P. Fagerstroem, Ph.D. Thesis, University of Toronto (1999), 
DESY Report DESY-THESIS-1999-039.

\bibitem{geant}
R. Brun et al., {\it GEANT3 }, Technical Report CERN DD/EE/84-1, CERN,
1987.


\bibitem{jetset} T. Sj\"ostrand, \CPC{82}{1994}{74}.

\bibitem{django} G. A. Sch\"uler and H. Spiesberger, {\it Proceedings of
the Workshop on Physics at HERA}, Volume~3, W. Buchm\"uller and G.
Ingelman (eds.), DESY (1991), p.~1419;\\
H. Spiesberger, DJANGOH, {\verb+http://www.desy.de/~hspiesb/djangoh.html+.}

\bibitem{lepto} G. Ingelman, A.\ Edin and J.\ Rathsman, \CPC{101}{1997}{108}.

\bibitem{igm} F.O.\ Dur\~aes, F.S.\ Navarra and G.\ Wilk,  
\PR{D 58}{1998}{094034}.

\bibitem{igm1} G.N. Fowler et al., \PR {C 40}{1989}{1219}.

\bibitem{earlydiffraction} ZEUS Collab., M. Derrick et al., \ZP{C 
68}{1995}{569}.


\bibitem{SNS1} H. Holtmann et al., Z. Phys. {\bf C 69}, 297 (1996).

\bibitem{whitmore} J. Whitmore et al., \PR{D 11}{1975}{3124}.


\bibitem{ganguli} S.N. Ganguli and D.P. Roy, Phys. Rep. {\bf 67}, 201
(1980).

\bibitem{batista} M. Batista and R.J.M. Covolan, Phys. Rev. {\bf D 59},
054006 (1999).

\bibitem{igm_pc} F.O. Dur\~aes, F.S. Navarra and G. Wilk, hep-ph/0209328 
(2002).

\bibitem{kolya_private} N.N. Nikolaev, private communication.

\bibitem{lps94} ZEUS Collab., J. Breitweg et al., \EP{C 2}{1998}{237}.

\bibitem{kolya_APP} N.N. Nikolaev, Acta Phys. Pol. {\bf B 29}, 2425 
(1998).



\bibitem{michiel} M. Botje,  \EP{C 14}{2000}{285}.


\end{thebibliography}
\end{document}